\documentclass[a4paper, 12 pt, twoside, reqno]{amsart}

\usepackage[english]{babel}
\usepackage[utf8]{inputenc}

\usepackage{setspace}
\usepackage[euler-digits,euler-hat-accent]{eulervm}
\usepackage{times}
\usepackage{amsmath}
\usepackage{amsfonts}
\usepackage{amssymb}
\usepackage{amsmath}
\usepackage{amsthm}
\usepackage{graphicx}
\usepackage{array}
\usepackage{color}
\usepackage{mathrsfs}
\usepackage{hyperref}
\usepackage{eucal}
\usepackage{esint} %%% for \fint
\usepackage{tikz}
\usepackage{upgreek}
\usepackage{enumitem}

\allowdisplaybreaks

\theoremstyle{plain}
\newtheorem{theorem}{Theorem}[section]
\newtheorem{corollary}[theorem]{Corollary}
\newtheorem{proposition}[theorem]{Proposition}
\newtheorem{lemma}[theorem]{Lemma}

\theoremstyle{definition}
\newtheorem{definition}[theorem]{Definition}

\theoremstyle{remark}
\newtheorem{remark}[theorem]{Remark} 
\newtheorem{example}[theorem]{Example}

\numberwithin{equation}{section}
\numberwithin{figure}{section}
\numberwithin{table}{section}

\newcommand{\R}{\mathbb{R}}
\newcommand{\N}{\mathbb{N}}
\newcommand{\C}{\mathbb{C}}

\newcommand{\s}[1]{\CMcal{#1}}
                  
\newcommand{\bb}[1]{\mathscr{#1}}
\newcommand{\rr}[1]{\mathfrak{#1}}
\newcommand{\n}[1]{\mathbb{#1}}

\newcommand{\bra}[1]{\langle#1|}
\newcommand{\ket}[1]{|#1\rangle}
     
\newcommand{\ketbra}[2]{|#1\rangle\langle#2|}

\newcommand{\expo}[1]{\,\mathrm{e}^{#1}\,}                 

\newcommand{\dd}{\,\mathrm{d}}
\newcommand{ \ii}{\,\mathrm{i}\,}

\newcommand{\virg}[1]{\lq\lq#1\rq\rq}                \newcommand{\ie}{\textsl{i.\,e.\,}}
\newcommand{\eg}{\textsl{e.\,g.\,}}
\newcommand{\cf}{\textsl{cf}.\,}

\DeclareMathOperator{\Tr}{Tr}

\hypersetup{
pdftoolbar=true,        % show Acrobat’s toolbar?
pdfmenubar=true,        % show Acrobat’s menu?
pdffitwindow=true,     % window fit to page when opened
pdfstartview=true,    % fits the width of the page to the window
pdftitle={NCG-Landau-II},    % title
pdfauthor={Giuseppe De Nittis},     % author
breaklinks=true, %permet le retour a la ligne dans les liens trop longs
colorlinks=true,       % false: boxed links; true: colored links
linkcolor=purple,         % color of internal links (change box color 
citecolor=teal, % color of links to bibliography
urlcolor=blue, %couleur des hyperliens
bookmarksopen=true, %si les signets Acrobat sont crees,
filecolor=magenta,      % color of file links
}

\begin{document}

\title[Dixmier trace and  the DOS of magnetic operators]{
Dixmier trace and  the DOS of magnetic operators}

\author[F. Belmonte]{Fabian Belmonte}

\address[F. Belmonte]{Departamento de Matemáticas, Universidad Cat\'olica del Norte, Antofagasta, Chile.}
\email{fbelmonte@ucn.cl}

\author[G. De~Nittis]{Giuseppe De Nittis}

\address[G. De~Nittis]{Facultad de Matemáticas \& Instituto de Física,
  Pontificia Universidad Católica de Chile,
  Santiago, Chile.}
\email{gidenittis@mat.uc.cl}

\vspace{2mm}

\date{\today}

\begin{abstract}
The main goal of this work is to provide two new formulas for
the computation of the trace per unit volume,
and consequently 
 the integrated density of states (IDOS), for magnetic operators. These formulas also permit the use of the Dixmier trace in the spectral analysis of  magnetic operators. The second of these formulas, named \emph{energy shell formula}, permits to approximate the IDOS by a finite sums
 of  averaged expectation values of the spectral projections.
\medskip

\end{abstract}

\maketitle

\tableofcontents

\section{Introduction}\label{sec:Intr0}
The main goal of this work is to provide two new formulas for
the computation of the trace per unit volume,
and consequently 
 the density of states (DOS), for magnetic operators. These formulas also permit to use  the Dixmier trace in the spectral analysis of  these operators.
In order to describe the new results we first  need to introduce some definitions and notations. We will use \cite{denittis-sandoval-00}, and references therein, as the source for this background material.

\medskip

The \emph{(dual) magnetic translations}\footnote{The name magnetic translations is common in the condensed matter community since the works of Zak \cite{zak1,zak2}. Mathematically, they are known as Weyl systems.}
on $L^2(\R^2)$ are the unitary operators 
defined by
\begin{equation}\label{eq:intro_000}
  \left(V(a)\psi\right)(x)\; =\; \expo{\ii\frac{x\wedge a}{2\ell^{2}}}\;\psi(x-a)\;, \qquad a \in \R^2\;,\quad \psi \in L^2(\R^2)\;
 \end{equation}
where $x\wedge a :=x_{1}a_{2} -
    x_{2}a_{1}$ for all $x=(x_1,x_2)$ and $a=(a_1,a_2)$.
A direct 
  computation shows that 
  \[
  \begin{aligned}
    V(a)V(b)\;&=\; \expo{\ii\frac{b\wedge a}{2\ell^{2}}}\; V(a+b)
  \end{aligned}
  \;,\qquad a,b \in \R^2\;.
\]
The parameter $\ell>0$ is known as \emph{magnetic length}. From a  physical point of view  it is proportional to $\beta^{-\frac{1}{2}}$ where $\beta>0$ is the strength of a constant magnetic field perpendicular to the plane $\R^2$. Therefore, $\ell \to \infty$ represents the \virg{singular} limit of a vanishing magnetic field 
\cite[Remark 2.2]{denittis-sandoval-00}.
Let $\bb{V}:=C^*(V(a),a\in\R^2)$ be the $C^*$-algebra generated by the unitaries $V(a)$. The \emph{magnetic von Neumann algebra} $\bb{M}$ (or magnetic algebra, for short) is by definition the {commutant} of $\bb{V}$  \cite[Proposition 2.18]{denittis-sandoval-00} \ie,
\begin{equation}\label{int_VNA}
\bb{M}\;:=\;\bb{V}'\:.
\end{equation}
The name magnetic algebra is justified by the fact the \emph{Landau Hamiltonian} 
\begin{equation}\label{eq:intro_LH}
  H_{\rm L}\;: =\; \frac{1}{2} \left(-\ii \ell\frac{\partial}{\partial x_{1}}  - \frac{1}{2 \ell} x_{2}\right)^2\;+\;
  \frac{1}{2} \left(-\ii \ell\frac{\partial}{\partial x_{2}}  + \frac{1}{2 \ell} x_{1}\right)^2\;,
\end{equation}
is affiliated to $\bb{M}$   \cite[Section 2.3]{denittis-sandoval-00}.

\medskip

The magnetic algebra $\bb{M}$ admits a canonical, faithful, semi-finite and normal (FSN) trace $\tau$, defined on the two-sided 
self-adjoint ideal $\bb{I}_\tau\subset \bb{M}$  \cite[Proposition 2.21]{denittis-sandoval-00}. We will provide a precise definition  
for $\bb{I}_\tau$ and  $\tau$ in Section \ref{sec:BG_mat}. For the purpose of this introduction it is important to point out that $\tau$ can be realized as the \emph{trace per unit  volume}. For that, let  $\Lambda_{n} \subseteq \R^2$ be
an
 increasing sequence of compact subsets such that $\Lambda_n\nearrow\R^2$ and which satisfies the \emph{F{\o}lner condition} (see \cite{greenleaf-69} for more details). 
 Let $\chi_{\Lambda_n}$ be the projection defined as the multiplication operator by the characteristic function of $\Lambda_n$.
 A bounded operator $S$ admits the trace per unit  volume (with respect  to the  F{\o}lner sequence $\Lambda_n$) if the limit
 \[
\s{T}_{\rm u.v.}(S)\;:=\;\lim_{n\to+\infty} \frac{1}{|\Lambda_n|}{\Tr}_{L^2(\R^2)}( \chi_{\Lambda_n}S \chi_{\Lambda_n} )\;
 \]
 exists. It follows that 
every $S\in\bb{I}_\tau$ admits the trace per unit  volume,  and 
\begin{equation}\label{eq:intro_001}
\tau(S)\;=\; \frac{\Omega_\ell}{2}\;\s{T}_{\rm u.v.}(S)\;, \qquad \Omega_\ell\;:=\;\pi(2\ell)^2 
\end{equation}
independently of the election of any particular  F{\o}lner  sequence \cite[Lemma 2.23 \& Remark 2.24]{denittis-sandoval-00}. It is interesting to notice that the constant $\Omega_\ell$ has the physical meaning of the area of the \emph{magnetic disk} of radius $2\ell$. 
For the aims of this introduction equation \eqref{eq:intro_001} can be used as the definition of $\tau$, although a more direct and intrinsic definition will be provided in Section \ref{sec:BG_mat}.
It is worth to point out that 
a crucial aspect for the existence of the thermodynamic limit defining $\s{T}_{\rm u.v.}$ is that the elements of $\bb{I}_\tau$ are left invariant by the action of $\R^2$ implemented by the magnetic translations \eqref{eq:intro_000}.

\medskip 

The trace per unit  volume plays a crucial role for the study of the spectral and thermodynamic properties of quantum systems. An important example is given by the DOS of a Hamiltonian as discussed in Section \ref{Sect_DOS}. For this reason it is important to have formulas that allow to calculate, or approximate, the trace per unit  volume.
One of the main contribution of this work is to provides two new formulas which allow  to  
compute $\tau$. The first of such formulas relates the computation of the trace to the estimate of a residue. To state this result we need
to introduce the family of operators
\begin{equation}\label{eq:res_Q}
 Q_{\lambda}^{-s}\;:=\;(Q\;+\;\lambda{\bf 1})^{-s}\;
\end{equation}
with $s>0$ and $\lambda >-1$ where %
\begin{equation}\label{eq:Harm_osc}
 Q\;:=\;-\ell^2\; \Delta\;+\;\frac{1}{4\ell^2}|x|^2
\end{equation}
is the two-dimensional isotropic \emph{harmonic oscillator} given by the sum of the Laplacian
 $\Delta:=\partial_{x_1}^2+\partial_{x_2}^2$ and  the  harmonic potential $|x|^2:=x_1^2+x_2^2$.
 Since the spectrum of $Q$ is $\sigma(Q)=\N=\{1,2,3,\ldots\}$, it follows that $Q+\lambda\bf{1}$ is invertible for every $\lambda >-1$. 
\begin{theorem}[Residue formula]\label{theo:1st-RF}
For every $S\in\bb{I}_\tau$ it holds true that
$$
\tau(S)\;=\;\lim_{x\to 0^+}x\;{\Tr}\left(Q_{\lambda}^{-(1+x)}S\right)\;.
$$
\end{theorem}
\medskip
\noindent
The proof of Theorem \ref{theo:1st-RF} requires several intermediate steps and will be presented in Section \ref{sect:pre-res-form}.

\medskip

For the second formula let us recall that the harmonic oscillator $Q$  is diagonalized by the  Laguerre
 basis $\psi_{n,m}$  defined in \eqref{eq:lag_pol}. More precisely, one has that
 \[
Q \psi_{n,m}\;=\;(n+m+1)\ \psi_{n,m},\qquad\quad\;  \forall\;(n,m)\in\N^2_0\;,
\]
where $\N_0:=\N\cup\{0\}$.
As a consequence  the   eigenspace of $Q$ associated to the eigenvalue $j\in\N$, also called $j$-th \emph{energy shell}, is spanned by the $\psi_{n,m}$ such that $n+m=j-1$ and has dimension (or degeneracy) $j$. Therefore, the quantity
\begin{equation}\label{eq:avr}
 w_j(S)\;:=\;\frac{1}{j}\sum_{n+m=j-1}\langle\psi_{n,m}, S\psi_{n,m}\rangle_{L^2}\;,\qquad  j\in \N
\end{equation}
represents the 
\emph{averaged expectation value} of the operator $S$ on the $j$-th {energy shell} of $Q$.
\begin{theorem}[Energy shell formula]\label{theo:2st-ES}
For every $S\in\bb{I}_\tau$ it holds true that
$$
\tau(S)\;=\; \lim_{N\to+\infty}\left(\frac{1}{\log(N)}\sum_{j=1}^{N}w_j(S)\right)\;.
 $$
\end{theorem}
\medskip
\noindent
The proof of Theorem \ref{theo:2st-ES} is postponed to Section \ref{sec:shell_F}.

 \medskip

 The trace per unit volume, or equivalently the trace $\tau$ in view of \eqref{eq:intro_001},
is the central object for the construction of the  \emph{integrated density of states} (IDOS)  $N_H$ associated to the self-adjoint  operator $H$  affiliated with $\bb{M}$. Albeit the precise description of $N_H$ (which requires also the assumption of the \emph{spectral regularity} for $H$)   will be provided in Definition \ref{def:idos}, we can anticipate that $N_H$ is a positive, non-decreasing and right-continuous function defined on $\R$ (Lemma \ref{lemma_idos}).
According to Definition \ref{def:dos}, the DOS of $H$ is the  \emph{Lebesgue–Stieltjes measure} associated with $N_H$
  and will be denoted with $\mu_{H}$.  Let $\epsilon_\infty:=\sup \sigma(H)$ the supremum of the spectrum of $H$ (with the convention that $\epsilon_\infty=+\infty$ when $H$ is unbounded from above) and $C_{\rm c}((-\infty,\epsilon_\infty))$  the space of  compactly supported continuous functions on the open interval 
$(-\infty,\epsilon_\infty)$. Then,
the following \emph{spectral formula}
\begin{equation}\label{eq:spec_form}
\tau\big(f(H)\big)\;=\;\frac{\Omega_\ell}{2}\;\int_{\R}\dd\mu_H(\epsilon)\; f(\epsilon) 
\end{equation}
 holds true for every $f\in C_{\rm c}((-\infty,\epsilon_\infty))$ (Proposition \ref{prop:spec_form}).
 The combination of  formula \eqref{eq:spec_form} with Theorem \ref{theo:2st-ES} provides the following result:
\begin{theorem}[Spectral energy shell formula]\label{theo:approx_for}
Let $H$ be a self-adjoint and spectrally regular operator affiliated to the magnetic algebra $\bb{M}$. For every $f\in C_{\rm c}((-\infty,\epsilon_\infty))$ the following asymptotic formula holds true
 $$
 \lim_{N\to+\infty}\left(\frac{1}{\log(N)}\sum_{j=1}^{N}w_j(f(H))\right)\;=\;\frac{\Omega_\ell}{2}\;\int_\R \dd\mu_H(\epsilon)\; f(\epsilon)\;.
 $$
\end{theorem}

\medskip

{In the same vein, by combining Theorem \ref{theo:2st-ES} with the definition of the IDOS $N_H$ (Definition \ref{def:dos}) one obtains the following approximated formula} 
\begin{equation}\label{form_approx}
N_H(\epsilon)\;=\;\lim_{N\to+\infty}\;
\frac{2}{\Omega_\ell\log(N)}\sum_{j=1}^{N}w_j(P_H(\epsilon))\;,\quad\text{as}\;\;N\to+\infty\;
\end{equation}
where $P_H(\epsilon)$ is the spectral projection of $H$ for the spectral interval $(-\infty,\epsilon]$.
 Formula \eqref{form_approx} seems to be new in the literature and can be useful for numerical computations. To some extent this formula plays for magnetic (continuous) operators a similar role that the \emph{local DOS} and the \emph{windowed DOS} play for tight-binding models \cite{loring-lu-watson-21}.

 \medskip
 
An important result obtained in \cite{denittis-sandoval-00} 
ensures that on  a certain $\ast$-subalgebra $\bb{L}^1$ of $\bb{I}_\tau$ (see Section \ref{sec:BG_mat} for the precise definition)
the trace $\tau$ can be expressed  in terms of the \emph{Dixmier trace} weighed\footnote{{It is  worth pointing out that the use of the 
\virg{weight} $Q_{\lambda}^{-1}$ in \eqref{eq:traXXX_III_XX}
has a precise justification in therms of the noncommutative geometry of the \emph{magnetic spectral triple} for the magnetic algebra constructed in \cite{denittis-sandoval-00}.
In such a geometric context it is  more appropriate to refer to $Q_{\lambda}^{-1}$ as the \virg{magnetic volume form}. 
 These aspects have been discussed in detail in \cite{denittis-sandoval-00}, and  we will not dwell further on this point.}} by $Q^{-1}_\lambda$. 
{We will assume here some
familiarity of the reader with the theory of the Dixmier trace and we provide in
Appendix \ref{sec:dix_tr_2} a short summary of the main properties of the Dixmier trace along with some useful references.}
For the moment let us recall that the natural domain of the Dixmier trace is an ideal of the compact operators denoted with $\rr{S}^{1^+}$ and called the \emph{Dixmier ideal}. The construction of the Dixmier trace requires the choice of a generalized scale-invariant limit $\omega$. However, there exists a relevant subset $\rr{S}^{1^+}_{\rm m}\subset \rr{S}^{1^+}$ where the computation of the Dixmier trace   is independent of the election of $\omega$. The elements of $\rr{S}^{1^+}_{\rm m}$ are called \emph{measurable} operators. The action of any Dixmier trace on  $\rr{S}^{1^+}_{\rm m}$ will be denoted simply with ${\Tr}_{{\rm Dix}}:\rr{S}^{1^+}_{\rm m}\to\C$.
  In \cite[Proposition 2.27]{denittis-sandoval-00} it has been proved that when 
 $S\in\bb{L}^1$ then 
  the  equality
\begin{equation}\label{eq:traXXX_III_XX}
\begin{aligned}
\tau(S) \;=\; {\Tr}_{\rm Dix}\big(T_S\big)\end{aligned}
\end{equation}
holds true 
for every
\begin{equation}\label{eq:traXXX_III_XX_bis}
T_S\;\in\;\left\{Q_{\lambda}^{-1}S,SQ_{\lambda}^{-1},Q_{\lambda}^{-\frac{1}{2}}SQ_{\lambda'}^{-\frac{1}{2}}\right\}\;\subset\;\rr{S}^{1^+}_{\rm m}\;,
\end{equation}
independently of  $\lambda,\lambda'>-1$.
In this case the spectral formula \eqref{eq:spec_form} can be reinterpreted as follows:
 \begin{theorem}\label{corol:main_dix_eq_03}
Let $H$ be a self-adjoint and $\bb{L}^1$-spectrally regular operator affiliated to the magnetic algebra $\bb{M}$. For every $f\in C_{\rm c}((-\infty,\epsilon_\infty))$ let
$$
T_{f(H)}\;\in\;\left\{Q_{\lambda}^{-1}f(H),f(H)Q_{\lambda}^{-1},Q_{\lambda}^{-\frac{1}{2}}f(H)Q_{\lambda'}^{-\frac{1}{2}}\right\}\;.
$$
Then $T_{f(H)}\in\rr{S}^{1^+}_{\rm m}$ and
it holds true that 
$$
{\Tr}_{\rm Dix}\big(T_{f(H)}\big)\;=\;\frac{\Omega_\ell}{2}\;\int_\R \dd\mu_H(\epsilon)\; f(\epsilon)\;
$$
independently of  $\lambda,\lambda'>-1$.
\end{theorem}
\medskip
\noindent
The proof of Theorem \ref{corol:main_dix_eq_03} follows directly form formula
  \eqref{eq:traXXX_III_XX} and Proposition \ref{prop:spec_form} applied to $\bb{L}^1$-spectrally regular operators. The latter property (described in Definition \ref{def:idos}) and  
  Lemma \ref{lemm_absorb}  ensure that $f(H)\in \bb{L}^1$, and in turn the applicability of the equation \eqref{eq:traXXX_III_XX}. The restriction imposed by the $\bb{L}^1$-spectral regularity could be removed if one could prove that equality \eqref{eq:traXXX_III_XX} holds for every $S\in\bb{I}_\tau$. This aspect will be discussed in Section \ref{sec:pos_ghen}.

 \medskip
 
Theorem \ref{corol:main_dix_eq_03} represents the \virg{magnetic version} of \cite[Theorem 1.1]{azamov-mcdonald-sukochev-zanin-19} in the special case $d=2$. Aside from the obvious similarity, there are few differences we would like to point out. Theorem 1.1 in \cite{azamov-mcdonald-sukochev-zanin-19}
works for Schr\"odinger operators of  type $-\Delta+V$ with  real-valued potentials $V\in L^\infty(\R)$, while Theorem \ref{corol:main_dix_eq_03} is valid for every ($\bb{L}^1$-spectrally regular) magnetic operator affiliated with the magnetic algebra $\bb{M}$ but without potentials. Perturbations by potentials will be considered in a future work
along the lines anticipated in Section \ref{sec:pos_ghen}. 
 Anyway, it is suggestive to compare the case of the \emph{free Laplacian} \cite[Example 1.2]{azamov-mcdonald-sukochev-zanin-19} with the case of the \emph{free Landau operator} in Example \ref{ex:free_lap}. There is also a second relevant difference. The \virg{weight} introduced in \cite[Theorem 1.1]{azamov-mcdonald-sukochev-zanin-19} is the multiplication operator by the function $(1+|x|^2)^{-1}$ which is evidently not compact. 
On the contrary, the \virg{weight} $Q_{\lambda}^{-1}$ used in Theorem \ref{corol:main_dix_eq_03} is a compact operator which is diagonalized on the basis of the  {generalized Laguerre functions} $\psi_{n,m}$ defined in \eqref{eq:lag_pol}.
The latter fact provides a significant computational advantage. In fact
it is the hidden reason behind the energy shell formulas of Theorem \ref{theo:2st-ES} and 
Theorem \ref{theo:approx_for}.

 \medskip
 
 \noindent
{\bf Structure of the paper.}
In {\bf Section~\ref{sec:BG_mat}} we will introduce the background material about magnetic operators necessary as the starting point for the  formulation of our main results.  
{\bf Section~\ref{sect:pre-res-form}}
contains the proof of the 
 {residue-type formula} anticipated in Theorem \ref{theo:1st-RF} and {\bf Section~\ref{sec:shell_F}} contains the  the proof of  the energy shell formula provided in Theorem \ref{theo:2st-ES}.
{\bf Section~\ref{Sect_DOS}} concerns with the construction 
of the IDOS and of the DOS and contains the proof of the spectral formula \eqref{eq:spec_form}, and consequently of 
Theorem \ref{corol:main_dix_eq_03}. 
Possible generalizations of our main results  along with the related open problems  are briefly discussed in 
{\bf Section~\ref{sec:pos_ghen}}.
 {\bf Appendix~\ref{sec:ser-von}} contains some useful results about the elements of the magnetic algebra $\bb{M}$
which are needed for the proof of Lemma \ref{lemm_absorb}.
 Finally, {\bf Appendix~\ref{sec:dix_tr_2}} contains a very short introduction to the Dixmier trace
 in order to make this  work self-contained.

 \medskip
 
 \noindent
{\bf Acknowledgements.}
GD's research is supported by the grant \emph{Fondecyt Regular - 1190204}. GD would like to cordially thank Massimo Moscolari  for his help in the construction of the proof of Theorem \ref{theo:1st-RF}.

%--------%

\section{Relevant aspects about magnetic operators}\label{sec:BG_mat}
The material presented in this preliminary section is borrowed from \cite{denittis-gomi-moscolari-19,denittis-sandoval-00}.
Consider  the Hilbert space $L^2(\R^2)$ and let $\{\psi_{n,m}\}\subset L^2(\R^2)$, with
$n,m\in\N_0$, be the orthonormal 
\emph{Laguerre
  basis} defined by 
\begin{equation}\label{eq:lag_pol}
\psi_{n,m}(x)\;:=\;\psi_0(x)\ \sqrt{\frac{n!}{m!}}\left[\frac{x_1+\ii x_2}{\ell\sqrt{2}}\right]^{m-n}L_{n}^{(m-n)}\left(\frac{|x|^2}{2\ell^2}\right)\; ,
\end{equation}
where
\[
  L_n^{(\alpha)}\left(\zeta\right)\;:=\;\sum_{j=0}^{n}\frac{(\alpha+n)(\alpha+n-1)\ldots(\alpha+j+1)}{j!(n-j)!}\left(-\zeta\right)^j\;,\quad\alpha,\zeta\in \R
\]
are the {generalized Laguerre polynomial} of degree $m$ (with the usual convention $0!=1$) and 
\begin{equation}\label{eq:herm1}
\psi_{0}(x)\;:=\;\frac{1}{\sqrt{2\pi}\ell}\ \expo{-\frac{|x|^2}{4\ell^2}}\;.
\end{equation}
From the definition it follows that $\psi_{0,0}=\psi_0$.

\medskip

Let us introduce the family $\{\Upsilon_{j\mapsto k}\;|\, (j,k)\in \N_0^2\}$ of \emph{transition operators}  on $L^2(\R^2)$ defined by 
\begin{equation}\label{eq:intro:basic_op}
\Upsilon_{j\mapsto k}\psi_{n,m}\;:=\;\delta_{j,n}\;\psi_{k,m}\;,\qquad k,j,n,m\in\N_0\;.
\end{equation}
A direct computation shows that  \cite[Proposition 2.10]{denittis-sandoval-00}
\begin{equation}\label{eq:rel_alg}
(\Upsilon_{j\mapsto k})^*\;=\;\Upsilon_{k\mapsto j}\;,\qquad \Upsilon_{j\mapsto k}\Upsilon_{m\mapsto n}\;=\;\delta_{j,n}\Upsilon_{m\mapsto k}
\end{equation}
for every $j,k,n,m\in \N_0^2$. 
The relations \eqref{eq:rel_alg} allow to introduce the $C^*$-algebra 
\begin{equation}\label{eq:equal_C-Ups}
\bb{C}\;=\;C^*(\Upsilon_{j\mapsto k},\;k,j\in\N_0)
\end{equation}
 generated inside the algebra of bounded operators
$\bb{B}(L^2(\R^2))$ by  the norm closure of polynomials in the generators $\Upsilon_{j\mapsto k}$.  It turns out that $\bb{C}$ is non-unital.
We will refer to  $\bb{C}$ as the \emph{magnetic $C^*$-algebra}. Such a name is justified by the fact that the Landau Hamiltonian
$H_{\rm L}$ defined in \eqref{eq:intro_LH} is affiliated with $\bb{C}$. More precisely, it turns out that the \emph{Landau projections} $\Pi_j:=\Upsilon_{j\mapsto j}$ are the spectral projections of $H_{\rm L}$ which provide the spectral representation
\begin{equation}\label{eq:LH_spec_res}
H_{L}\; =\; \sum_{j\in\N_0}\lambda_j\;\Pi_j\;,\qquad \lambda_j\;:=\;\left(j+\frac{1}{2}\right)\;.
\end{equation}
The eigenvalue $\lambda_j$ is known as the $j$-th \emph{Landau level}.

\medskip

The  magnetic  algebra $\bb{M}$ defined by \eqref{int_VNA} coincides with the enveloping von Neumann algebra of $\bb{C}$  \ie,
 $
\bb{M}=\bb{C}''
$ 
where on the right-hand side one has the bicommutant  of  $\bb{C}$ \cite[Proposition 2.18]{denittis-sandoval-00}. 
Besides $\bb{C}$, there are other interesting subspaces contained in  $\bb{M}$. Let us introduce the following family of spaces
\begin{equation}\label{eq:exp_op}
\bb{L}^p\;:=\;\left.\left\{A\;:=\;\sum_{(j,k)\in\N_0^2}a_{j,k}\Upsilon_{j\mapsto k}\;\right|\;\{a_{j,k}\}\in \ell^p(\N_0^2)\right\}\;,
\end{equation}
where $\ell^p(\N_0^2)$ are the usual  spaces of $p$-summable sequences on $\N_0^2$. 
Every  $\bb{L}^p$, obtained as the closure of the polynomials in the generators $\Upsilon_{j\mapsto k}$ with respect the associated $\ell^p$-norms, turns out to be a Banach space. 
One has that  \cite[Proposition 2.17]{denittis-sandoval-00}
$$
\bb{L}^1\;\subset\;\bb{I}_\tau\;\subset\;\bb{L}^2\;\subset\;\bb{C}\;\subset\;\bb{M}\;,
$$
where
$$
\bb{I}_\tau\;:=\;\left\{S=AB\;|\; A,B\in\bb{L}^2 \right\}\;\equiv\;\left(\bb{L}^2\right)^2\;.
$$
All these subspaces are dense in $\bb{C}$ with respect to the operator norm, and in $\bb{M}$ with respect to the weak or strong topology. Both $\bb{L}^2$, and consequently $\bb{I}_\tau$, are 
self-adjoint two-sided ideals of $\bb{M}$ \cite[Proposition 2.18]{denittis-sandoval-00}.

\medskip

The space  $\bb{L}^2$ admits a special characterization 
in terms of integral kernel operators \cite[Section 2.4]{denittis-sandoval-00}. Indeed, it turns out that $A\in \bb{L}^2$ if and only if there is a $f_A\in L^2(\R^2)$ such that
\begin{equation}\label{eq:bg1}
(A\varphi)(x)\;=\;\frac{1}{2\pi \ell^{2}} \int_{\mathbb{R}^{2}}\dd y\; f_A(y-x)\;\expo{\ii\frac{x\wedge y}{2\ell^{2}}}\;\varphi(y)\;,
\;
\quad \forall\;\varphi \in L^2(\R^2)\;.
\end{equation}
The relation between the integral kernel $f_A$ and the sequence 
$\{a_{j,k}\}\in \ell^2(\N_0^2)$ which identifies the expansion of $A$ in the basis $\Upsilon_{j\mapsto k}$ is given by
\begin{equation}\label{eq:bg2}
f_A\;=\;\sqrt{2\pi}\ell\sum_{(j,k)\in\N_0^2}(-1)^{j-k}a_{j,k}\;\psi_{k,j}\;
\end{equation}
and the norm bound $\sqrt{2\pi}\ell\;\|A\|\leqslant\|f_A\|_{L^2}$ holds true.

\medskip

The space $\bb{L}^1$ is a Banach $\ast$-algebra with respect to the $\ell^1$-norm \cite[Lemma B.1]{denittis-sandoval-00}.
It is not  an ideal of $\bb{M}$ but fulfills the following \virg{absorption} property:
\begin{lemma}\label{lemm_absorb}
Let $A_1,A_2\in \bb{L}^1$ and $T\in\bb{M}$. Then $A_1TA_2\in\bb{L}^1$.
\end{lemma}
\noindent
The proof of this result is postponed to Appendix \ref{sec:ser-von}.

\medskip
As discussed in \cite[Section 2.6]{denittis-sandoval-00}, 
the von Neumann algebra $\bb{M}$ admits 
a canonical, faithful, semi-finite and normal (FSN) trace
 $\tau$ defined
on the ideal $\bb{I}_\tau$, and uniquely specified by the prescription
\begin{equation}\label{eq:trac_prod}
\tau(A^*B)\;:=\;\frac{1}{2\pi\ell^2}\langle f_A,f_B\rangle_{L^2}\;,\qquad \forall\; A,B\in\bb{L}^2
\end{equation}
where $\langle\;,\;\rangle_{L^2}$ is the usual scalar product in $L^2(\R^2)$ and $f_A,f_B\in L^2(\R^2)$ are the integral kernels of $A$ and $B$ respectively, as given by \eqref{eq:bg2}. {By using the orthonormality of the Laguerre functions and the expansions of $A$ and $B$ in terms of the operators $\Upsilon_{j\mapsto k}$ and the $\ell^2$-sequences $\{a_{j,k}\}$ and $\{b_{j,k}\}$ respectively, one gets from \eqref{eq:trac_prod} the following formula
\begin{equation}\label{eq:trac_prod_X}
\tau(A^*B)\;:=\;\sum_{(j,k)\in\N_0^2}\overline{a_{j,k}}\;b_{j,k}\;.
\end{equation}
Let $S:=A^*B$ be a generic element of $\bb{I}_\tau$. Then, a direct computation shows that $S$ can be expanded as
 \begin{equation}\label{eq:tra_0_00}
S\;=\;\sum_{(j,k)\in\N_0^2}s_{j,k}\Upsilon_{j\mapsto k}\qquad \text{with}\qquad s_{j,k}\;:=\;\sum_{n\in\N_0}\overline{a_{k,n}}\;b_{j,n}\;,
\end{equation}
and a comparison with \eqref{eq:trac_prod} provides
 \begin{equation}\label{eq:tra_0}
 \tau(S)\;=\;\sum_{k\in\N_0}s_{k,k}\;:=\;
\lim_{N\to+\infty}\left(\sum_{k=0}^Ns_{k,k}\right)\;,\qquad\forall\; S\in\bb{I}_\tau\;.
\end{equation}
Finally, let us recall that every $S\in \bb{I}_\tau$ as an integral kernel 
 of type \eqref{eq:bg2} which satisfies $f_S\in L^2(\R^2)\cap C_0(\R^2)$, where  $C_0(\R^2)$ is the space of continuous functions which vanish at infinity. Therefore, it makes sense to evaluate $f_S$ pointwise and one can prove that  $ \tau(S)= f_S(0)$ \cite[Corollary 2.22]{denittis-sandoval-00}
 }

\medskip

The harmonic oscillator $Q$ defined by \eqref{eq:Harm_osc} 
 admits the spectral decomposition
\begin{equation}\label{eq:spec_avr_Q}
Q\;=\;\sum_{(n,m)\in\N^2_0}(n+m+1)\Pi_n P_m
\end{equation}
which involves the Landau projections $\Pi_n$ and the \emph{transverse projections}  $P_m$ defined by
\begin{equation}\label{eq:transv_project}
P_m\psi_{j,k}\;:=\;\delta_{m,k}\;\psi_{j,k}\;,\qquad k,j,m\in\N_0\;.
\end{equation}
It is important to notice that
$
\Pi_n P_m= P_m\Pi_n$ for every $(n,m)\in\N^2_0
$.
It turns out that 
$$
 Q_{\lambda}^{-s}\;=\;\sum_{(n,m)\in\N^2_0}\frac{1}{(n+m+1+\lambda)^{s}}\;\Pi_n P_m
$$
provides the spectral representation of the operator defined by 
\eqref{eq:res_Q}. In particular the last equation shows that $Q_{\lambda}^{-s}$ is a well defined compact operator for every 
$s>0$ and $\lambda>-1$.

\medskip

{
The relation between the family  $Q_{\lambda}^{-s}$ and the Dixmier trace has been investigated in \cite[Lemma B.4 \& Lemma B.5]{denittis-gomi-moscolari-19} and in \cite[Corollary 2.26 \& Lemma 3.10]{denittis-sandoval-00}. Let $\rr{S}^{p}$ be the $p$-th
\emph{Schatten ideal}, $\rr{S}^{p^+}$  the 
\emph{Ma\u{c}aev ideal} of order $p^+$ and $\rr{S}^{1^+}_{\rm m}\subset \rr{S}^{1^+}$ the space of \emph{measurable} operators (\cf Appendix \ref{sec:dix_tr_2}).
It results that $Q_{\lambda}^{-s}\in\rr{S}^{1}$ for every $s>2$ and $\lambda>-1$. Moreover, $Q_{\lambda}^{-2}\in \rr{S}^{1^+}_{\rm m}$  and
$$
{\Tr}_{\rm Dix}\big(Q_{\lambda}^{-2}\big)\;=\;\frac{1}{2}
$$
independently of $\lambda>-1$. When $Q_{\lambda}^{-1}$ is multiplied by an element in  $\bb{L}^1$ one obtains the equality \eqref{eq:traXXX_III_XX} \ie, 
\begin{equation}\label{eq:dix_ser_dig}
{\Tr}_{\rm Dix}\big(T_S\big)\;=\;\sum_{k\in\N_0}s_{k,k}
\end{equation}
where $T_S$ is one of the operators in \eqref{eq:traXXX_III_XX_bis} and $\{s_{j,k}\}$ are the coefficients of $S\in\bb{L}^1$.
}

%-----------%
\section{The residue formula}\label{sect:pre-res-form}
In this section we will show that the trace $\tau$ on $\bb{I}_\tau$ can be computed via the {residue-type} formula 
anticipated  in Theorem \ref{theo:1st-RF}.

\medskip

Let us start by observing that the Laguerre basis induces the isomorphism of Hilbert spaces $L^2(\R^2)\simeq\ell^2(\N_0^2)$
via the unitary map $\s{U}:\psi_{n,m}\mapsto \ket{n,m}$. Here,  the Dirac notation\footnote{{We will assume a certain familiarity of  the reader with the Dirac notation. Let us just recall that $\ketbra{n',m'}{n,m}$ denotes the rank-one operator with initial space spanned by $\ket{n,m}$ y target space  spanned by $\ket{n',m'}$. }} $\ket{n,m}$ for the canonical basis of $\ell^2(\N_0^2)$ is used.
Every operator $T$ acting on  $\ell^2(\N_0^2)$ can be represented as
$$
T\;=\;\sum_{(n',m')\in\N_0^2}\sum_{(n,m)\in\N_0^2}\kappa_T[(n',m'),(n,m)]\;\ketbra{n',m'}{n,m}
$$
for a given function $\kappa_T:\N_0^2\times \N_0^2\to\C$ {defined through the matrix elements
\begin{equation}\label{eq:gen_int_ker}
\kappa_T[(n',m'),(n,m)]\;:=\;\bra{n',m'}T\ket{n,m}\;.
\end{equation}
}
Every vector $\phi\in\ell^2(\N_0^2)$ can be expanded as $\phi:=\sum_{(n,m)\in\N_0^2}\phi(n,m)\;\ket{n,m}$ and the explicit computation 
$$
(T\phi)(n,m)\;=\;\sum_{(n',m')\in\N_0^2}\kappa_T[(n,m),(n',m')]\;\phi(n',m')\;,
$$
shows that $\kappa_T$ is the \emph{integral kernel} of $T$.

\medskip

{
The integral kernel associate to the transition operator $\Upsilon_{j\mapsto k}$ is given by
$$
\kappa_{\Upsilon_{j\mapsto k}}[(n,m),(n',m')]\;:=\;\bra{n,m}\s{U}\Upsilon_{j\mapsto k}\s{U}^{-1}\ket{n',m'}\;=\;\delta_{n,k}\;\delta_{n',j}\;\delta_{m,m'}\;.
$$
More in general, let $A\in\bb{L}^p$ with associated coefficients $\{a_{n,n'}\}\in\ell^p(\N_0^2)$ according to \eqref{eq:exp_op}. Then, the associated integral kernel is given by
\begin{equation}\label{eq:ker_01}
\kappa_A[(n,m),(n',m')]\;:=\;\bra{n,m}\s{U}A\s{U}^{-1}\ket{n',m'}\;=\;a_{n',n}\delta_{m,m'}\;.
\end{equation}
In the same way the integral kernel associate to $Q_{\lambda}^{-s}$ is given by
\begin{equation}\label{eq:ker_01.2}
\begin{aligned}
\kappa_{Q_{\lambda}^{-s}}[(n,m),(n',m')]\;:&=\;\bra{n,m}\s{U}Q_{\lambda}^{-s}\s{U}^{-1}\ket{n',m'}\\&=\;\frac{1}{(n+m+1+\lambda)^s}\;\delta_{n,n'}\delta_{m,m'}\;.
\end{aligned}
\end{equation}
}

\medskip

We will need the family of auxiliary operators
\begin{equation}\label{eq:Mr}
M^r\;:=\;\sum_{m\in \N_0}(m+1)^r\; P_m
\end{equation}
where $r\in\R$ is a real parameter and $P_m$ are the transverse projections \eqref{eq:transv_project}. 
{For $r\leqslant 0$ equation \eqref{eq:Mr} defines a bounded operator, while for $r>0$
one has an unbounded operator densely defined on the finite linear combination of the Laguerre functions. From its very definition it is immediate to recognize that $M^r$ commutes with $Q_{\lambda}^{-s}$ since they share the same family of spectral projections.
On the other hand, $M^r$ also commute
 with every element of the magnetic algebra $\bb{M}$ since it belongs to (if $r\leqslant 0$), or is affiliated\footnote{{In the unbounded case $r>0$ the resolvents of $M^r$ are elements of $\bb{M}'$ and thus commute with $\bb{M}$. This is enough to ensure that every $A\in \bb{M}$ sends the vectors in the domain of $M^r$ into the domain of $M^r$ itself,  and the equation $AM^r-M^r A=0$ is initially well defined  on the domain of $M^r$ (\cf the proof of \cite[Lemma 2.19]{denittis-sandoval-00}). By a standard continuity argument one then extends the commutation relation to the full Hilbert space.}}  with (if $r>0$) the commutant $\bb{M}'$. 
The integral kernel of $M^r$ is given by
\begin{equation}\label{eq:ker_02}
\kappa_{M^r}[(n,m),(n',m')]\;:=\;\bra{n,m}\s{U}M^r\s{U}^{-1}\ket{n',m'}\;=\;(m+1)^r\;\delta_{n,n'}\;\delta_{m,m'}\;.
\end{equation}
}

\medskip

Let  $\rr{S}^2$ be the ideal of Hilbert-Schmidt operators (the $2$-nd {Schatten ideal}) on $L^2(\R^2)$. We are now in position to prove some  preliminary results.
\begin{lemma}\label{lemma:aux_1}
Let $A\in \bb{L}^2$. Then  
$$
M^{-r}A\;=\;AM^{-r}\;\in\; \rr{S}^2
$$ 
for every $r>\frac{1}{2}$.
\end{lemma}
\proof
First of all a bounded operator $B$ on $L^2(\R^2)$ is Hilbert-Schmidt if and only if $\s{U}B\s{U}^{-1}$ is Hilbert-Schmidt
as operator on $\ell^2(\N_0^2)$. Secondly, the property of being 
Hilbert-Schmidt on $\ell^2(\N_0^2)$ can be checked by showing that the associated integral kernel is in $\ell^2(\N_0^2\times \N_0^2)$ \cite[Theorem VI.23]{reed-simon-I}. By using 
 formulas \eqref{eq:ker_01} and \eqref{eq:ker_02} on can compute explicitly the integral kernel of $\s{U}M^{-r}A\s{U}^{-1}$ which is given by
$$
\kappa_{M^{-r}A}[(n,m),(n',m')]\;:=\;\;a_{n',n}\;(m+1)^{-r}\;\delta_{m,m'}\;.
$$
Therefore
$$
\begin{aligned}
\big\|\kappa_{M^{-r}A}\big\|_{\ell^2}^2\;&=\;\sum_{(n,n',m)\in \N_0^3}|a_{n',n}|^2\;(m+1)^{-2r}\\
&=\;\frac{\|f_A\|_{L^2}^2}{{2\pi}\ell^2}\;\sum_{m\in \N_0}(m+1)^{-2r}
\end{aligned}
$$
and $f_A\in L^2(\R^2)$ is the convolution kernel of $A$ as given by
\eqref{eq:bg1} and \eqref{eq:bg2}. It is evident that the sum on the right-hand side converges whenever
$r>\frac{1}{2}$. 
\qed

\begin{lemma}\label{lemma:aux_2}
Let $A\in \bb{L}^2$. Then  
$$
M^{r}Q_{\lambda}^{-s}A\;=\;Q_{\lambda}^{-s}M^{r}A\;=\;Q_{\lambda}^{-s}AM^{r}\;\in\; \rr{S}^2
$$ 
 for every $r\geqslant0$ and $s>0$ such that $s-r>\frac{1}{2}$.
\end{lemma}
\proof
We will use the same  strategy of  the proof of Lemma \ref{lemma:aux_1}. The integral kernel of 
$\s{U}M^{r}Q_{\lambda}^{-s}A\s{U}^{-1}$ can be computed from 
\eqref{eq:ker_01}, \eqref{eq:ker_01.2} and \eqref{eq:ker_02}
and is given by
\begin{equation}\label{eq:ker_A}
\kappa_{M^{r}Q_{\lambda}^{-s}A}[(n,m),(n',m')]\;:=\;\frac{(m+1)^r}{(n+m+1+\lambda)^s}\;a_{n',n}\;\delta_{m,m'}\;.
\end{equation}
Therefore
$$
\begin{aligned}
\big\|\kappa_{M^{r}Q_{\lambda}^{-s}A}\big\|_{\ell^2}^2\;&=\;\sum_{(n,n',m)\in \N_0^3}\frac{(m+1)^{2r}}{(n+m+1+\lambda)^{2s}}\;|a_{n',n}|^2\\
&\leqslant\;\sum_{(n,n',m)\in \N_0^3}\frac{(m+1)^{2r}}{(m+1+\lambda)^{2s}}\;|a_{n',n}|^2\\
&=\;\frac{\|f_A\|_{L^2}^2}{{2\pi}\ell^2}\;\sum_{m\in \N_0}\frac{(m+1)^{2r}}{(m+1+\lambda)^{2s}}
\end{aligned}
$$
and the last series converges whenever $2(s-r)>1$.\qed

\medskip

Let  $\rr{S}^1$ be the ideal of trace-class operators 
(the $1$-st {Schatten ideal}) on $L^2(\R^2)$. It is known that 
 for a given pair $A,B\in \rr{S}^2$ of Hilbert-Schmidt operators, the product $AB\in \rr{S}^1$ is trace-class.
\begin{corollary}\label{cor:aux_101}
Let $A,B\in \bb{L}^2$. Then  
$$
\left\{Q_{\lambda}^{-s}AB,\; BQ_{\lambda}^{-s}A,\; ABQ_{\lambda}^{-s}\right\}\;\subset\; \rr{S}^1
$$ 
for every $s>1$. Moreover,
$$
{\Tr}\big(Q_{\lambda}^{-s}AB\big)\;=\;{\Tr}\big(BQ_{\lambda}^{-s}A\big)\;=\;{\Tr}\big(AB Q_{\lambda}^{-s}\big)\;.
$$
\end{corollary}
\proof
Let $\frac{1}{2}<r<\frac{1}{2}+(s-1)$.
The identity
$$
Q_{\lambda}^{-s}AB\;=\;\left(Q_{\lambda}^{-s}AM^r\right)\left(M^{-r}B\right)
$$
along with  Lemma \ref{lemma:aux_1} and Lemma \ref{lemma:aux_2} shows that $Q_{\lambda}^{-s}AB$ is the product of two Hilbert-Schmidt operators, hence it is trace-class. The  remaining cases 
can be proved in a similar way.
The equality of the traces is guaranteed by \cite[Corollary 3.8]{simon-05}.
\qed

\begin{corollary}\label{cor:aux_1.2}
Let $S\in \bb{I}_\tau$. Then  
$$
Q_{\lambda}^{-\frac{s}{2}}SQ_{\lambda}^{-\frac{s}{2}}\;\in\; \rr{S}^1
$$ 
for every $s>1$ and
$$
{\Tr}\left(Q_{\lambda}^{-\frac{s}{2}}SQ_{\lambda}^{-\frac{s}{2}}\right)\;=\;{\Tr}\big(Q_{\lambda}^{-s}S\big)\;=\;{\Tr}\big(S Q_{\lambda}^{-s}\big)\;.
$$
\end{corollary}
\proof
Let $S=AB$ with $A,B\in\bb{L}^2$. Since Lemma \ref{lemma:aux_2} ensures that $Q_{\lambda}^{-\frac{s}{2}}A$ and $BQ_{\lambda}^{-\frac{s}{2}}$ are both Hilbert-Schmidt, their product is trace-class. The equality of the traces is guaranteed by \cite[Corollary 3.8]{simon-05}.
\qed

\medskip

Corollary \ref{cor:aux_101} and Corollary \ref{cor:aux_1.2} establish that for every
$S\in\bb{I}_\tau$ the products
$Q_{\lambda}^{-s}S$, $SQ_{\lambda}^{-s}$ and $Q_{\lambda}^{-\frac{s}{2}}SQ_{\lambda}^{-\frac{s}{2}}$ are  trace-class operators, whenever  $s>1$. This allows to associate to every $S\in\bb{I}_\tau$
the function $\theta_S:(0,+\infty)\to\C$ defined equivalently by one of the following equalities:
\begin{equation}\label{def_theta}
\begin{aligned}
\theta_S(x)\;:&=\;{\Tr}\left(Q_{\lambda}^{-(1+x)}S\right)\;=\;{\Tr}\left(SQ_{\lambda}^{-(1+x)}\right)\\
&=\;{\Tr}\left(Q_{\lambda}^{-\frac{1+x}{2}}SQ_{\lambda}^{-\frac{1+x}{2}}\right)\;.
\end{aligned}
\end{equation}

\proof[Proof of Theorem \ref{theo:1st-RF}.]
It is sufficient to prove the claim for positive elements in $\bb{I}_\tau$. In fact, every   $S\in\bb{I}_\tau$ is a linear combination of at most four non-negative elements and the 
functions $S\mapsto \theta_S(x)$ and $S\mapsto\tau(S)$  are linear.
Then, let us assume $S\geqslant 0$. This implies that 
$s_{n,n}\geqslant 0$ for all $n\in\N_0$ where $\{s_{n,m}\}\subset\ell^2(\N_0^2)$ are the coefficients associated with the expansion of  $S$ as in \eqref{eq:exp_op}.
The computation of the trace of $Q_{\lambda}^{-s}S$ can be  performed by integrating \virg{along the diagonal} the integral kernel of $\s{U}Q_{\lambda}^{-(1+x)}S\s{U}^{-1}$. This kernel can be computed by putting $r=0$ in equation \eqref{eq:ker_A} and  one gets
\begin{equation}\label{eq:ker_B}
\kappa_{Q_{\lambda}^{-(1+x)}S}[(n,m),(n',m')]\;:=\;\frac{1}{(n+m+1+\lambda)^{1+x}}\;s_{n',n}\;\delta_{m,m'}\;.
\end{equation}
 Therefore one obtains
$$
\begin{aligned}
\theta_S(x)\;&=\;\sum_{(n,m)\in\N_0^2}\kappa_{Q_{\lambda}^{-(1+x)}S}[(n,m),(n,m)]\\
&=\;\sum_{(n,m)\in\N_0^2}\frac{s_{n,n}}{(n+m+1+\lambda)^{1+x}}\;.
\end{aligned}
$$
The latter is a convergent series with positive terms (hence, absolutely convergent) and therefore it can be rearranged as
\begin{equation}\label{eq:ser_thet}
\theta_S(x)\;=\;\sum_{n\in\N_0}s_{n,n}\;\zeta(1+x,n+1+\lambda)
\end{equation}
where
$$
\zeta(t,q)\;:=\;\sum_{m\in\N_0}\frac{1}{(m+q)^t}\;,\qquad t>1\;,\quad q>0
$$
is the \emph{Hurwitz zeta function} \cite[Chapter 64]{oldham-myland-spanier-09}. 
It is known that $\zeta(t,q)$ has a simple pole with residue $1$ in $t=1$ , \ie
$$
\lim_{t\to1^+}g_q(t)\;=\;1
$$
where $g_q(t):=(t-1)\zeta(t,q)$. 
Moreover,  one has that 
$$
\lim_{t\to+\infty}g_q(t)\;=\;0\;,\qquad \forall\; q>1\;.
$$
Then, $g_q$ is a bounded continuous function on $(1,+\infty)$
and 
$$
\|g_q\|_{\infty}\;:=\;\sup_{t\in (1,+\infty)}|g_{q}(t)|\;<\;+\infty\;,\qquad \forall\; q>1\;.
$$
From $\zeta(t,q)<\zeta(t,q+\delta)$ when $\delta>0$, one infers that 
$$
\|g_{q+\delta}\|_{\infty}\;\leqslant\;\|g_q\|_{\infty}\;.
$$
We can use the expression \eqref{eq:ser_thet} to write
$$
x\theta_S(x)\;=\;s_{0,0}\;g_{1+\lambda}(1+x)\;+\;\left(\sum_{n\in\N}s_{n,n}\;g_{n+1+\lambda}(1+x)\right)\;
$$
{where therm of order zero has been isolated to consider possible $\lambda>-1$}. 
The  series in the right is uniformly 
convergent since
$$
\left\|\sum_{n\in\N}s_{n,n}\;g_{n+1+\lambda}(1+x)\right\|_\infty
\;\leqslant\;\|g_{2+\lambda}\|_{\infty}\; \sum_{n\in\N}s_{n,n}\;=\;\|g_{2+\lambda}\|_{\infty}\;(\tau(S)-s_{0,0})
$$
in view of the relation \eqref{eq:tra_0}.
Therefore, one gets
$$
\begin{aligned}
\lim_{x\to 0^+}x\;\theta_S(x)\;=\;\sum_{n\in\N_0}s_{n,n}\;\lim_{t\to1^+}(t-1)\zeta(t,n+1+\lambda)\;=\;\sum_{n\in\N_0}s_{n,n}\;=\;\tau(S)
\end{aligned}
$$
where the last equality is provided again by \eqref{eq:tra_0}.
\qed

\medskip

The next formula may be useful in some application.

\begin{corollary}\label{cor:redef_theta}
Let $A,B\in\bb{L}^2$. Then
$$
\lim_{x\to 0^+}x\;{\Tr}\left(BQ_{\lambda}^{-(1+x)}A^*\right)\;=\;\frac{1}{2\pi\ell^2}\langle f_A,f_B\rangle_{L^2}
$$
where $f_A,f_B$ are the kernels associated to $A$ and $B$ respectively by \eqref{eq:bg2}.
\end{corollary}
\proof
Let $S:=A^*B\in\bb{I}_\tau$. By combining  Corollary \ref{cor:aux_101} with the definition \eqref{def_theta} one gets
$$
{\Tr}\left(BQ_{\lambda}^{-(1+x)}A^*\right)\;=\;\theta_S(x)\;.
$$
Therefore, in view of Theorem \ref{theo:1st-RF}, the limit in the claim equals $\tau(S)=\tau(A^*B)$. Finally, it is enough to apply equation \eqref{eq:trac_prod}.
\qed

%----------%
\section{The energy shell formula}\label{sec:shell_F}
In this section we will build the proof of Theorem \ref{theo:2st-ES}. We will start with a preliminary result valid for elements in  $\bb{L}^2$.
\begin{lemma}\label{lemm:pre_approx_inter}
Let 
$$
S\;=\; \sum_{(j,k)\in\N_0^2}s_{j,k}\;\Upsilon_{j\mapsto k}\;\in\;\bb{L}^2\;.
$$
Then, it holds true that
\begin{equation}\label{eq:sumN->}
 \lim_{N\to+\infty}\left(\frac{1}{\log(N)}\sum_{j=1}^{N}w_j(S)\right)\;=\; \lim_{N\to+\infty}\left(\sum_{n=0}^Ns_{n,n}\right)\;
 \end{equation}
 where $w_j(S)$ is defined by \eqref{eq:avr}.
\end{lemma}
\proof
 A direct computation
based on \eqref{eq:intro:basic_op}, and the orthonormality of the Laguerre basis provide
$$
\langle\psi_{n,m}, S\psi_{n,m}\rangle_{L^2}\;=\;s_{n,n}\;,\qquad \forall\;(n,m)\in\N_0^2 \;.
$$
Therefore, in view of \eqref{eq:avr} one gets that 
$$
w_j(S)\;=\;\frac{1}{j}\sum_{\substack{(n,m)\in\N_0^2 \\  n+m=j-1}}s_{n,n}\;=\;\frac{1}{j}\sum_{n=0}^{j-1}s_{n,n}\;.
$$
A rearrangement of the  double finite sum provides
$$
\begin{aligned}
\sum_{j=1}^{N}w_j(S)\;&=\;s_{0,0}\;+\;\frac{1}{2}(s_{0,0}+s_{1,1})\;+\;\ldots\;+\;\frac{1}{N}\sum_{n=0}^{N-1}s_{n,n}\\
&=\;s_{0,0}\rr{h}_{N}\;+\;s_{1,1}(\rr{h}_N-1)\;+\;\ldots\;\frac{1}{N}s_{N-1,N-1}\\
&=\;\sum_{n=0}^{N-1}(\rr{h}_N-\rr{h}_n)s_{n,n}
\end{aligned}
$$
where $\rr{h}_n:=\sum_{k=1}^nk^{-1}$ is the $k$-th \emph{harmonic number} with the convention  $\rr{h}_0=0$.
By a  Cauchy-Schwarz argument one observes that 
$$
\left|\sum_{n=0}^{N-1}\rr{h}_ns_{n,n}\right|^2\;\leqslant\;\left(\sum_{n=0}^{N-1}\rr{h}_n^2\right)\left(\sum_{n=0}^{N-1}|s_{n,n}|^2\right)\;<\;\frac{\pi^2}{6}\; C_s
$$
where the first factor in the constant on the right is given by infinite sum  of the $\rr{h}_n^2$ and the second factor is $C_s:=\sum_{n\in\N_0}|s_{n,n}|^2$ which is finite in view of the fact that 
$\{s_{j,k}\}\in \ell^2(\N_0^2)$. This implies that 
$$
\lim_{N\to+\infty}\left(\frac{1}{\log(N)}\sum_{n=0}^{N-1}\rr{h}_ns_{n,n}\right)\;=\;0
$$
and in turn
$$
\lim_{N\to+\infty}\left(\frac{1}{\log(N)}\sum_{j=1}^{N}w_j(S)\right)\;=\;\lim_{N\to+\infty}\left(\frac{\rr{h}_N}{\log(N)}\sum_{n=0}^{N-1}s_{n,n}\right).
$$
Since $\rr{h}_N\sim \log(N)$ when $N\to+\infty$ one gets the desired result. 
\qed

\medskip

{Equality \eqref{eq:sumN->} must be understood in the sense that the two limits have the same behavior. In particular, it says that one of the two limits converges to a finite value if and only if the other also converges to the same value. However, the two limits can be both divergent or indefinite. For instance,}
as long as $S$ belongs to $\bb{L}^2$ there is no guarantee that the series on the right-hand side of \eqref{eq:sumN->} converges. On the other hand, the convergence is evidently guaranteed if one restricts to elements in $\bb{L}^1$. However, there is a less restrictive condition which guarantee the convergence. Let $A,B\in \bb{L}^2$ and consider $S=A^*B$. Then, {by using \eqref{eq:tra_0_00}} one gets 
$$
\left|\sum_{n\in\N_0}s_{n,n}\right|^2\;=\;\left|\sum_{(n,k)\in\N_0^2}\overline{a_{n,k}}\;b_{n,k}\right|^2\;\leqslant\; \left\|\{a_{i,j}\}\right\|_{\ell^2}^2\;\left\|\{b_{i,j}\}\right\|_{\ell^2}^2\;<\;+\infty
$$
in view of the Cauchy–Schwarz inequality. At this point we have all the ingredients to prove the 
energy shell formula.

\proof[Proof of Theorem \ref{theo:2st-ES}.]
Let {$S\in \bb{I}_\tau$. Then, according to \eqref{eq:tra_0} one has that 
$$
\lim_{N\to+\infty}\left(\sum_{n=0}^Ns_{n,n}\right)\;=\;\sum_{n\in\N_0}s_{n,n}\;=\;\tau(S)
$$
and this concludes the proof.} 
\qed

\medskip

In the case $S\in \bb{L}^1$ we can identify the trace $\tau$ with the Dixmier trace according to 
\eqref{eq:traXXX_III_XX}.
In such a case one has the following result.
\begin{corollary}\label{prop:pre_approx}
Let $S\in\bb{L}^1$ and $T_S$ one of the elements of the set 
\eqref{eq:traXXX_III_XX_bis}.
Then, it holds true that
$$
{\Tr}_{\rm Dix}\big(T_S\big)\;=\; \lim_{N\to+\infty}\left(\frac{1}{\log(N)}\sum_{j=1}^{N}w_j(S)\right)\;,
 $$
independently of   $\lambda,\lambda'>-1$.
\end{corollary}

\medskip

Corollary \ref{prop:pre_approx} is reminiscent of the formula for the computation of the Dixmier trace of \emph{modulated operators} \cite[Section 11.2]{lord-sukochev-zanin-12}. This aspect will be explored a little more in Section \ref{sec:pos_ghen}.

%------%

\section{IDOS and DOS}\label{Sect_DOS}
In this section we will 
introduce the notions of {integrated density of states} (IDOS) and {density of states} (DOS) from  a slighted generalized point of view, and we will
provide the relation between  DOS  and Dixmier trace for magnetic operators. For a general exposition on the theory of the DOS we will refer to the modern monograph \cite{veselic-08}, as well as to the classical textbooks \cite{carmona-lacroix-90} or \cite{pastur-figotin-92}.

\medskip

In this section $\bb{M}$ will denote a von Neumann algebra of bounded operators acting on the Hilbert space $\s{H}$ and endowed with a {normal}, {faithful} and {semi-finite}   trace $\tau$ defined
on the ideal $\bb{I}_\tau$. When necessary, $\bb{M}$ will be interpreted as the magnetic algebra introduced in Section \ref{sec:BG_mat}. 
A (non necessary bounded) self-adjoint operator
 $H$ acting on $\s{H}$ is \emph{affiliated} to  $\bb{M}$ 
if for every borelian subset $\Sigma\subseteq\sigma(H)$ of the spectrum of $H$ the associated
 spectral projection $\chi_{\Sigma}(H)$ belongs to $\bb{M}$.
 For every given \virg{energy} $\epsilon\in\R$ let $P_H(\epsilon):=\chi_{(-\infty,\epsilon]}(H)$ be the  spectral projection  associated with the set $(-\infty,\epsilon]$. Let $\epsilon_\infty:=\sup \sigma(H)$ be the supremum of the spectrum of $H$ with the convention that $\epsilon_\infty=+\infty$ when $H$ is unbounded from above. 
\begin{definition}[IDOS and spectral regularity]\label{def:idos}
Let $\bb{Y}\subseteq\bb{I}_\tau$ be a $\ast$-subalgebra with the following  absorption property
\begin{equation}\label{eq:con_abs}
Y_1 T Y_2\;\in\; \bb{Y}
\end{equation}
for every $Y_1,Y_2\in \bb{Y}$ and every $T\in\bb{M}$.
Let
$H$ be a self-adjoint operator 
affiliated with   $\bb{M}$. 
We will say that $H$ is $\bb{Y}$-\emph{spectrally regular}\footnote{When $\bb{Y}=\bb{I}_\tau$ we will refer simply to \emph{spectral regularity}  instead of $\bb{I}_\tau$-spectral regularity.} if 
$P_H(\epsilon)\in\bb{Y}$ for every $\epsilon<\epsilon_\infty$. In this case the function
$$
N_H(\epsilon)\;:=\;C\;\tau\left(P_H(\epsilon)\right)
$$
will be called the  \emph{integrated density of states} (IDOS) of $H$. The positive constant $C>0$ plays the role of a \virg{scale factor} for the IDOS.  
\end{definition}
\begin{remark}%\label{rk:no-inv}
Definition \ref{def:idos} is mainly meant for operators bounded from below. {For instance, in the special case of the magnetic algebra, the operator $-H_L$, where $H_L$ denotes the Landau Hamiltonian defined in \eqref{eq:intro_LH}, can not be spectrally regular.} The strict inequality in the condition $\epsilon<\epsilon_\infty$ serves to exclude the identity ${\bf 1}=P_H(\epsilon_\infty)$ which cannot be in the ideal of definition of the semi-finite trace $\tau$. When
$\bb{Y}$ is a subideal of $\bb{I}_\tau$ then condition \eqref{eq:con_abs}
is automatically
satisfied. In the specific case of the magnetic algebra 
$\bb{M}$ introduced in Section \ref{sec:BG_mat}
the scale factor for the IDOS is $C=(2\pi\ell^2)^{-1}$.
 \hfill $\blacktriangleleft$
\end{remark}

The main properties of $N_H$ are described below.
\begin{lemma}\label{lemma_idos}
For every self-adjoint and spectrally regular operator $H$ affiliated to $\bb{M}$ the function 
$N_H:(-\infty,\epsilon_\infty)\to\R$ is   positive,  non-decreasing and right-continuous. Moreover
$$
\begin{aligned}
N_H(-\infty)\;&:=\;\lim_{\epsilon\to-\infty}N_H(\epsilon)\;=\;0\;,\\
N_H(+\infty)\;&:=\;\lim_{\epsilon\to \epsilon_\infty}N_H(\epsilon)\;=\;+\infty\;.\\
\end{aligned}
$$
\end{lemma}
 \proof
The positivity follows from the positivity of $\tau$. Let $\epsilon'\leqslant \epsilon<\epsilon_\infty$ and consider the identity
 $P_H(\epsilon)=P_H(\epsilon')+\Delta_H(\epsilon',\epsilon')$ with 
 $\Delta_H(\epsilon',\epsilon):=\chi_{(\epsilon',\epsilon]}(H)$. 
 Therefore, the linearity and the positivity of $\tau$ imply
 $$
 N_H(\epsilon)\;=\;N_H(\epsilon')\;+\;C\;\tau\left(\Delta_H(\epsilon',\epsilon')\right)\;\geqslant\;N_H(\epsilon')\;
 $$
proving that $N_H$ is non-decreasing.
 Let $\delta>0$ (sufficiently small) and consider
$P_H(\epsilon+\delta)-P_H(\epsilon)=\Delta_H(\epsilon,\epsilon+\delta)=\chi_{(\epsilon,\epsilon+\delta]}(H)$. Observe that
$\lim_{\delta\to 0^+}\chi_{(\epsilon,\epsilon+\delta]}(x)=0$ for every $x\in\R$, namely the characteristic function $\chi_{(\epsilon,\epsilon+\delta]}$ converges point-wise to $0$.
Therefore, in view of the Borel functional calculus \cite[Theorem VIII.5 (d)]{reed-simon-I}
one gets that 
$$
{\rm s}-\lim_{\delta\to0^+}P_H(\epsilon+\delta)\;=\;P_H(\epsilon)\;,
$$
namely the family of projections  $P_H(\epsilon)$ is strongly right-continuous. The trace $\tau$ is normal, meaning that it is 
ultra-weakly continuous. Therefore, in view of \cite[Part I, Chap. 3, Theorem 1 (ii)]{dixmier-81} $\tau$ is equivalently ultra-strongly continuous and strongly continuous when restricted to the 
ball of  operators with  unitary norm. As a consequence the map $\epsilon\mapsto \tau\left(P_H(\epsilon)\right)$, and in turn the function $N_H$, are right-continuous on $\R$.
Since $\lim_{\epsilon\to-\infty}P_H(\epsilon)=0$ strongly, one obtains from the right-continuity that $N_H(-\infty)=0$.
On the other hand,  by observing that $\lim_{\epsilon\to\epsilon_\infty}P_H(\epsilon)={\bf 1}$ strongly,
and ${\bf 1}\in \bb{M}\setminus\bb{I}_\tau$ in view of the fact that $\tau$ is semi-finite,
one infers that $N_H(+\infty)=+\infty$.
 \qed
 
 \medskip
 
The passage from the IDOS to the DOS requires the use of the \emph{Lebesgue-Stieltjes measure} (\cf \cite[Sect. 15]{halmos-74} or \cite[Sect. 1.5]{folland-99}). 
\begin{definition}[Density of states]\label{def:dos}
Let $H$ be a self-adjoint and {spectrally regular} operator 
affiliated to   $\bb{M}$. Then, the   \emph{density of states} (DOS) of $H$ is by definition the Lebesgue-Stieltjes measure $\mu_H$ on $(-\infty,\epsilon_\infty)$ induced by $N_H$.
\end{definition}

{
\begin{remark}%\label{rk:no-inv}
By definition $\mu_H$ is the unique Borel measure on $\R$ defined by 
$$
\mu_H\big((\epsilon_1,\epsilon_2]\big)\;:=\;N_H(\epsilon_2)\;-\; N_H(\epsilon_1)
$$
via the  Carath\'eodory's extension. 
It follows from its very definition that
$$
N_H(\epsilon)\;=\:\int_{-\infty}^{\epsilon}\dd\mu_H(\epsilon')\;=\;\int_{-\infty}^{\epsilon}\dd\epsilon'\;\rho_H(\epsilon')
$$ 
where the second equality makes sense when $\mu_H$ is absolutely continuous with respect to the Lebesgue measure and
$\rho_H$ is its \emph{Radon-Nikodym derivative}. Sometimes  $\rho_H$  is referred as the  density of states of $H$ in the physical literature.
 \hfill $\blacktriangleleft$
\end{remark}
}

Let  $C_{\rm c}((-\infty,\epsilon_\infty))$ be the space of  compactly supported continuous functions on the open interval 
$(-\infty,\epsilon_\infty)$. By functional calculus $f(H)\in\bb{M}$.  We are now in position to prove the following relevant result:
\begin{proposition}[Spectral formula]\label{prop:spec_form}
Let $H$ be a self-adjoint and $\bb{Y}$-{spectrally regular} operator 
affiliated to  $\bb{M}$. Let 
$f\in C_{\rm c}((-\infty,\epsilon_\infty))$. Then 
$f(H)\in \bb{Y}$ and 
$$
\tau\big(f(H)\big)\;=\;\frac{1}{C}\;\int_{-\infty}^{\epsilon_\infty}\dd\mu_H(\epsilon')\; f(\epsilon')\;.
$$
\end{proposition}
\proof{
Since $f$ is compactly supported  there are $\epsilon_m<\epsilon_M<\epsilon_\infty$ such that the support of $f$ is contained into $[\epsilon_m,\epsilon_M]$. Therefore 
$f=\chi_{(-\infty,\epsilon_M]}f\chi_{(-\infty,\epsilon_M]}$ and by functional calculus this implies $f(H)=P_H(\epsilon_M)f(H)P_H(\epsilon_M)$ with $P_H(\epsilon_M)\in \bb{Y}$ by hypothesis.
Since $\bb{Y}$ meets the property \eqref{eq:con_abs}
one has that $f(H)\in \bb{Y}$.
The function $f$ can be approximated point-wise  (indeed uniformly) by a sequence of simple function $f_n$ \cite[Theorem 2.10]{folland-99}
such that $|f_n|\leqslant |f|\leqslant f_{\rm max}\chi_{(-\infty,\epsilon_M]}$ where 
$f_{\rm max}:=\|f\|_\infty$. Since $f$ is continuous and compactly supported, hence uniformly continuous, it turns out that the approximants $f_n$ can be constructed as Riemann partitions. For every $n\in N$ there is a $\delta_n$ such that  $|f(\epsilon)-f(\epsilon')|<2^{-n}$ whenever $\epsilon,\epsilon'\in [\epsilon_m,\epsilon_M]$ and $|\epsilon-\epsilon'|<\delta_n$. Fix a partition $\epsilon_m=:\epsilon_0<\ldots<\epsilon_k<\epsilon_{k+1}<\ldots<\epsilon_{N(n)}:=\epsilon_M$ such that $\epsilon_{k+1}-\epsilon_{k}<\delta_n$
and consider the  step function
$$
f_n(\epsilon)\;=\;\sum_{k=0}^{N(n)-1}f_n^k\;\chi_{(\epsilon_k,\epsilon_{k+1}]}(\epsilon)\;,\qquad -\infty\;<\;\epsilon\;<\;\epsilon_\infty
$$ 
where    $f_n^k:= f(\epsilon_{k+1})$. Let $\epsilon\in(\epsilon_k,\epsilon_{k+1}]$. Then
$$
|f(\epsilon)-f_n(\epsilon)|\;=\;|f(\epsilon)-f_n^k|\;=\;|f(\epsilon)-
 f(\epsilon_{k+1})|\;<\;2^{-n}
$$
and this shows that $f_n\to f$ point-wise.}
By observing that 
$\chi_{(\epsilon_k,\epsilon_{k+1}]}(H)=P_H(\epsilon_{k+1})-P_H(\epsilon_{k})$ and using the linearity of $\tau$ and the definitions of $N_H$ and $\mu_H$ one gets that
$$
\begin{aligned}
C\;\tau\big(\chi_{(\epsilon_k,\epsilon_{k+1}]}(H)\big)\;&=\;N_H(\epsilon_{k+1})\;-\; N_H(\epsilon_k)\\
&=\;\int_{-\infty}^{\epsilon_\infty}\dd\mu_H(\epsilon')\; \chi_{(\epsilon_k,\epsilon_{k+1}]}(\epsilon')
\end{aligned}
$$
and in turn
$$
\begin{aligned}
\tau\big(f_n(H)\big)\;
&=\;\frac{1}{C}\;\sum_{k=0}^nf_n^k\int_{-\infty}^{\epsilon_\infty}\dd\mu_H(\epsilon')\; \chi_{(\epsilon_k,\epsilon_{k+1}]}(\epsilon')\;.
\end{aligned}
$$
Therefore,  passing to the limit $n\to+\infty$ the right-hand side converge to the  Riemann-Lebesgue-Stieltjeg of $f$ with respect to the measure $\mu_H$. On the other hand the sequence $f_n(H)$ is equibounded by $f_{\rm max}$ and converges strongly to $f(H)$ in view of   the Borel functional calculus \cite[Theorem VIII.5 (d)]{reed-simon-I}. Since on bounded sequences the strong convergence implies the ultra-weak convergence, and  recalling that $\tau$ is normal, hence ultra-weakly continuous, one obtains that
$$
\begin{aligned}
\tau\big(f(H)\big)\;&=\;\lim_{n\to+\infty}\tau\big(f_n(\epsilon)\big)\\
&=\;\frac{1}{C}\;\lim_{n\to+\infty}\sum_{k=0}^nf_n^k\int_{-\infty}^{\epsilon_\infty}\dd\mu_H(\epsilon')\; \chi_{(\epsilon_k,\epsilon_{k+1}]}(\epsilon')\\
&=\;\frac{1}{C}\;\int_{-\infty}^{\epsilon_\infty}\dd\mu_H(\epsilon')\; f(\epsilon')\;.
\end{aligned}
$$
This concludes the proof.
\qed

\begin{remark}\label{rk:proof-Th1}
Equation  \eqref{eq:spec_form} follows from
 Proposition \ref{prop:spec_form} applied to the case of the magnetic algebra in which $C=2\Omega_\ell^{-1}$. Let us point out that in \eqref{eq:spec_form} we used the usual convention
 of thinking to the spectral measure $\mu_H$ as (trivially) extended on the complete real axis by the prescription $\mu_H(\R\setminus(-\infty,\epsilon_\infty])=0$.
\hfill $\blacktriangleleft$
\end{remark}

\medskip

\begin{example}[The Landau Hamiltonian]\label{ex:free_lap}
 It is worth applying the results of this section to the Landau Hamiltonian $H_L$ defined by \eqref{eq:intro_LH}. From the spectral representation of the Landau Hamiltonian \eqref{eq:LH_spec_res} one infers  that
 $P_{H_L}(\epsilon)=\sum_{j\in\N_0}\Theta(\epsilon-\lambda_j)\Pi_j$,
where $\Theta$ is the Heaviside step function, $\Pi_j$ is the $j$-th Landau projection  and $\lambda_j:=j+\frac{1}{2}$ is the $j$-th \emph{Landau level} \eqref{eq:LH_spec_res}. 
Since $\Pi_j\in\bb{L}^1$ for every $j\in\N_0$, one gets that 
$P_{H_L}(\epsilon)\in\bb{L}^1$ for every $\epsilon\in\R$. Therefore, $H_L$ is $\bb{L}^1$-spectrally regular. From $\tau(\Pi_j)=1$
 \cite[eq. (2.22)]{denittis-sandoval-00} one recovers the well known formula for the IDOS of the Landau Hamiltonian \cite[Appendix B]{nakamura-01}
$$
N_{H_L}(\epsilon)\;=\;\frac{1}{2\pi\ell^2}\;\sum_{j\in\N_0}\Theta(\epsilon-\lambda_j)\;.
$$ 
 The associated DOS can be represented by 
 $$
 \mu_{H_L}(\epsilon)\;:=\;\frac{\dd\epsilon}{2\pi\ell^2}\;\sum_{j\in\N_0}\delta(\epsilon-\lambda_j)
 $$
 as a sum of Dirac measures concentrated at the Landau levels.
The application of Theorem \ref{corol:main_dix_eq_03}
provides
$$
{\rm Tr}_{\rm Dix}\big(Q_{\lambda}^{-1}f(H_L)\big)\;=\;\sum_{j\in\N_0} f(\lambda_j)
$$ 
for every $f\in C_{\rm c}(\R)$.
  \hfill $\blacktriangleleft$
\end{example}

%-----------------%

\section{Some open question}\label{sec:pos_ghen}
There are two directions in which the main results presented in Section \ref{sec:Intr0} can be generalized.

\medskip

The first generalization concerns the fact that our proofs of Theorem \ref{corol:main_dix_eq_03} requires the assumption of  $\bb{L}^1$-spectral regularity. This  is a consequence of the fact that   formula
  \eqref{eq:traXXX_III_XX}  has been established 
only for elements $S\in \bb{L}^1$ \cite[Proposition 2.27]{denittis-sandoval-00}, or on a slightly bigger domain 
according to \cite[Theorem 2.28]{denittis-sandoval-00}.
However, it is our belief that formula
  \eqref{eq:traXXX_III_XX} should work for every element in the ideal $\bb{I}_\tau$ which is the natural domain of definition of the trace $\tau$   \cite[Remark 2.29]{denittis-sandoval-00}. This point is still an open conjecture which, once confirmed, would allow to extend Theorem \ref{corol:main_dix_eq_03} to every spectrally regular Hamiltonia $H$. In this work we have not been able to prove this conjecture. However, the residue formula 
  in Theorem \ref{theo:1st-RF}
    provides a small step toward the solution of this problem.
In fact our residue formula is very reminiscent of the \emph{Tauberian criterion} (Theorem \ref{theo:zeta_connes})
\begin{equation}\label{eq;taub}
\lim_{x\to 0^+}x\;{\Tr}\left(T^{1+x}\right)\;=\;{\Tr}_{{\rm Dix}}(T)
\end{equation}
which allows to compute the Dixmier trace of certain $T\in\rr{S}^{1^+}$ as a residue.
If one can prove the equality of the two residues in 
Theorem \ref{theo:1st-RF} and in \eqref{eq;taub} with $T=Q^{-1}_\lambda S$, then one would immediately obtain the proof of our conjecture. Also the energy shell formula in Theorem \ref{theo:2st-ES} is reminiscent of the fact that $\tau$ should be interpreted as a Dixmier trace  on all its domain of definition. In fact, by denoting with $\phi_r$ the orthonormal basis of eigenvectors of $Q^{-s}_0$ ($s>1$) ordered according to the decreasing sequence of the related eigenvalues (counting the multiplicity) one has that  
$$
\sum_{j=1}^{N}w_j(S)\;=\;\sum_{r=0}^{d_N}\langle\phi_{r}, T_S\phi_{r}\rangle_{L^2}
$$
where $d_N=\frac{1}{2}N(N-1)-1$ and $T_S$ is one of the operators in \eqref{eq:traXXX_III_XX_bis} with $\lambda=0$. 
For every $N\in\N$ let $M(N):=[\sqrt{2(N+1)}]$ where the $[\;\cdot\;]$ denotes the integer part.
From one hand one has that
$$
M(N)^2-M(N)\;\leqslant\;M(N)^2\;=\;\left[\sqrt{2(N+1)}\right]^2\;\leqslant\;2(N+1)
$$
which implies $d_{M(N)}\leqslant N$.
On the other hand
$$
(M(N)+1)^2+(M(N)+1)\;\geqslant\;(M(N)+1)^2\;=\;\left(\left[\sqrt{2(N+1)}\right]+1\right)^2\;\geqslant\;2(N+1)
$$
which implies $N\leqslant d_{M(N)+2}$.
Then
$$
\sum_{j=1}^{M(N)}w_j(S)\;\leqslant\;\sum_{r=0}^{N}\langle\phi_{r}, T_{S}\phi_{r}\rangle_{L^2}\;\leqslant\;\sum_{j=1}^{M(N)+2}w_j(S)
$$
and in turn
$$
\begin{aligned}
\frac{a_N}{\log(M(N))}\sum_{j=1}^{M(N)}w_j(S)\;&\leqslant\;\frac{1}{\log(N+1)}\sum_{r=0}^{N}\langle\phi_{r}, T_{S}\phi_{r}\rangle_{L^2}\\
&\leqslant\;\frac{b_{N}}{\log(M(N)+2)}\sum_{j=1}^{M(N)+2}w_j(S)
\end{aligned}
$$
where 
$$
a_N\;:=\;\frac{\log(M(N))}{\log(N+1)}\;,\qquad b_N\;:=\;\frac{\log(M(N)+2)}{\log(N+1)}\;.
$$ 
By observing that  
$$
\lim_{N\to+\infty}a_N\;=\;\lim_{N\to+\infty}b_N\;=\;\frac{1}{2}\;
$$
and using Theorem \ref{theo:2st-ES} one gets
\begin{equation}\label{eq:aks_01}
\lim_{N\to+\infty}\left(\frac{1}{\log(N+1)}\sum_{r=0}^{N}\langle\phi_{r}, T_{S}\phi_{r}\rangle_{L^2}\right)\;=\;\frac{1}{2}\;\tau(S)\;.
\end{equation}
The series on the left-hand side  is very reminiscent of the formula for the computation of the Dixmier trace of modulated operators \cite[Corollary 11.2.4 (c)]{lord-sukochev-zanin-12}. 
However, although this is a strong indication, we have not (yet) been able to adapt the theory of modulated operators \cite[Section 11.2]{lord-sukochev-zanin-12} to operators $T_S$ with $S\in\bb{I}_\tau$.

\medskip

The second type of generalization concerns the 
introduction of perturbations by (random)  electrostatic potentials. From a mathematical point of view this consists in replacing the magnetic $C^*$-algebra $\bb{C}$, which is the twisted group $C^*$-algebra of $\R^2$,
with the twisted $C^*$-crossed product generated by the  action $\rr{t}$ of $\R^2$ on the hull of the potentials $\Omega$ (which is a compact and Hausdorff space). In a more concrete way this 
$C^*$-crossed product can be thought of as the collection of
$C^*$-algebras $\bb{C}_\omega\subset\bb{B}(L^2(\R^2))$ parametrized by $\omega\in\Omega$, where every $\bb{C}_\omega$ is polynomially generated by
elementary operators of the type $M_g(\omega)\Upsilon_{j\mapsto k}$. Here  $\Upsilon_{j\mapsto k}$ are the 
transition operators defined in \eqref{eq:intro:basic_op}, 
$g\in C(\Omega)$ and $M_g(\omega)$ is the multiplication operator defined by 
$$
\big(M_g(\omega)\phi\big)(x)\;=\;g(\rr{t}_x(\omega))\;\phi(x)\;,\qquad x\in\R^2
$$
for every $\phi\in L^2(\R^2)$. It is reasonable to expect that
the content of Theorem \ref{corol:main_dix_eq_03} and Theorem \ref{theo:approx_for} could be extended to the case of the  
$C^*$-crossed product $\{\bb{C}_\omega\}_{\omega\in\Omega}$ provided that a certain 
averaging procedure with respect to the
ergodic probability measure $\n{P}$ on $\Omega$ is introduced. 
This problem is the object of current investigations.

%-----------%

\appendix

%-----------%

\section{Some properties  the magnetic algebra}
\label{sec:ser-von}
In \cite[Theorem A.1]{denittis-sandoval-00} it has been proved that every element $T\in\bb{M}$ of the magnetic algebra acts on $L^2(\R^2)$ as a twisted convolution operator with an  integral kernel given by a suitable tempered distribution $\Psi_T\in S'(\R^2)$. As proved in \cite{gracia-varilly-88} (just after the proof of Theorem 6) every tempered distribution admits an expansion in terms of the 
Laguerre basis\footnote{This result is similar to the $N$-representation theorem \cite[Theorem V.14]{reed-simon-I}  for $S'(\R^2)$ with the only  difference  that the Hermite basis  is replaced by the Laguerre basis.} $\psi_{k,j}$. 
Since every
transition operator $\Upsilon_{j\mapsto k}$ is nothing more than the twisted convolution with integral kernel $\psi_{k,j}$ (see the proof of \cite[Proposition 2.10]{denittis-sandoval-00}) it follows that every $T\in\bb{M}$ admits a series representation of the form
\begin{equation}\label{eq:exp_T}
T\;=\;\sum_{(j,k)\in\N_0^2}t_{j,k}\Upsilon_{j\mapsto k}\;
\end{equation}
{where the convergence of the series is meant in the strong (eq. weak) topology.}
The characterization of the behavior of the series of the coefficients $t_{j,k}$ is generally a difficult task and large part of the papers \cite{gracia-varilly-88,gracia-varilly-88-II} is devoted to this question. Here we will provide only a quite weak property.
\begin{lemma}\label{lem_bound_T}
Let $T\in\bb{M}$ and $t_{j,k}$ the coefficients   in the expansion \eqref{eq:exp_T}. 
Then,  $|t_{n,k}|\leqslant \|T\|$ for every $n,m\in\N_0$.
\end{lemma}
\proof
Since the elements of the Laguerre basis are normalized, one has that $\|T \psi_{n,m}\|_{L^2}\leqslant \|T\|$ for every $n,m\in\N_0$. By using the series representation \eqref{eq:exp_T} and the relation \eqref{eq:intro:basic_op} one gets
$$
T\psi_{n,m}\;=\;\sum_{(j,k)\in\N_0^2}t_{j,k}(\delta_{j,n}\;\psi_{k,m})\;=\;\sum_{k\in\N_0}t_{n,k}\;\psi_{k,m}
$$
and in turn
$$
\|T \psi_{n,m}\|_{L^2}^2\;=\;\sum_{k\in\N_0}|t_{n,k}|^2\; \leqslant\; \|T\|^2
$$
for every $n\in\N_0$. As a consequence none of the coefficients $|t_{n,k}|$ can exceed $\|T\|$.
\qed

\medskip

Lemma \ref{lem_bound_T} enters in the proof of Lemma \ref{lemm_absorb}.

\proof[Proof of Lemma \ref{lemm_absorb}]
Let 
$$
A_i\;=\;\sum_{(j,k)\in\N_0^2}a^{(i)}_{j,k}\Upsilon_{j\mapsto k}\;,\qquad i=1,2
$$
be two elements in $\bb{L}^1$ and  $T\in\bb{M}$. By using the 
the series representation \eqref{eq:exp_T} one gets
$$
A_1TA_2\;=\;\sum_{(r,s)\in\N_0^2}\sum_{(j,k)\in\N_0^2}\sum_{(p,q)\in\N_0^2}a^{(1)}_{r,s}t_{j,k}a^{(2)}_{p,q}(\Upsilon_{r\mapsto s}\Upsilon_{j\mapsto k}\Upsilon_{p\mapsto q})
$$
and since
$$
\Upsilon_{r\mapsto s}\Upsilon_{j\mapsto k}\Upsilon_{p\mapsto q}\;=\;\delta_{j,q}\delta_{r,k}\Upsilon_{p\mapsto s}
$$
one ends with
$$
A_1TA_2\;=\;\sum_{(p,s)\in\N_0^2}\kappa_{p,s}\Upsilon_{p\mapsto s}
$$
where
$$
\kappa_{p,s}\;:=\;\sum_{(r,q)\in\N_0^2}a^{(1)}_{r,s}t_{q,r}a^{(2)}_{p,q}\;.
$$
By invoking Lemma \ref{lem_bound_T} one has that
$$
\begin{aligned}
\sum_{(p,s)\in\N_0^2}|\kappa_{p,s}|\;&\leqslant\;
\sum_{(p,s)\in\N_0^2}\sum_{(r,q)\in\N_0^2}\left|a^{(1)}_{r,s}t_{q,r}a^{(2)}_{p,q}\right|\\
&\leqslant\;
\|T\|\;\sum_{(p,s)\in\N_0^2}\sum_{(r,q)\in\N_0^2}\left|a^{(1)}_{r,s}\right|\; \left|a^{(2)}_{p,q}\right|\\
&=\;
\|T\|\;\left\|\{a^{(1)}_{r,s}\}\right\|_{\ell^1}\;\left\|\{a^{(2)}_{p,q}\}\right\|_{\ell^1}\;\leqslant\;+\infty\;.
\end{aligned}
$$
Therefore, the coefficients $\kappa_{p,s}$ are in $\ell^1(\N_0^2)$ and as a consequence $A_1TA_2\in \bb{L}^1$.
\qed

\section{A primer on Dixmier trace}\label{sec:dix_tr_2}
 There are several standard references
 for the theory of the Dixmier trace, see \eg \cite[Chap.~4, Sect.~2]{connes-94}, \cite[Appendix A]{connes-moscovici-95}, \cite[Sect.~7.5 \& App.~7.C]{gracia-varilly-figueroa-01}, \cite[Chapter 6]{lord-sukochev-zanin-12}, \cite{alberti-matthes-02}.
  Here, we will recall only the basic facts concerning the Dixmier trace.
Let $\s{H}$ be a separable Hilbert space. The \emph{singular values} $\mu_n(T)$ of the compact operator $T\in\bb{K}(\s{H})$    are, by definition,   the eigenvalues of  $|T|:=\sqrt{T^*T}$.
By convention  the singular values will be listed in decreasing order, repeated according to their multiplicity, \ie
\[
\mu_0(T)\;\geqslant\;\mu_1(T)\;\geqslant\;\ldots\;\geqslant\;\mu_n(T)\;\geqslant\;\mu_{n+1}(T)\;\geqslant\;\ldots\;\geqslant\;0\;.
\]
In particular   $\mu_0(T) =\|\,|T|\,\|=\|T\|$.
Let
\begin{equation}\label{eq:partial_sigma}
\sigma_N^p(T)\;:=\;\sum_{n=0}^{N-1}\mu_n(T)^p\;,\qquad p\in[1,+\infty)\;.
\end{equation}
A compact operator $T$ is in the $p$-th
\emph{Schatten ideal} $\rr{S}^p$, if and only if,
 $$
 \|T\|_p^p\;:=\;\lim_{N\to\infty}\ \sigma_N^p(T)\;<\;+\infty\;.
 $$  Accordingly, $\rr{S}^1$ is the ideal
 of  trace-class operators. Let
\begin{equation}\label{eq:partial_gamma}
\gamma_N(T)\;:=\;\frac{\sigma_N^1(T)}{\log(N)}\;=\;\frac{1}{\log(N)}\sum_{n=0}^{N-1}\mu_n(T)\;,\qquad N>1\;.
\end{equation}
A compact operator $T$ is in the \emph{Dixmier ideal} $\rr{S}^{1^+}$ if its \emph{(Calder\'on)} norm
\begin{equation}\label{eq:clad_norm}
\lVert T\rVert_{1^+}\;:=\;\sup_{N>1}\ \gamma_N(T)\;<\;+\infty
\end{equation}
is finite. It turns out that $\rr{S}^{1^+}$ is a two-sided self-adjoint ideal  which is closed with respect to the norm~\eqref{eq:clad_norm} (but not
with respect to the operator norm). The set of operators such that
$\lim_{N\to\infty}\ \gamma_N(T)=0$ forms {an ideal} inside $\rr{S}^{1^+}$ denoted with $\rr{S}^{1^+}_0$. The latter coincides with the closure for the norm \eqref{eq:clad_norm} of the ideal of finite-rank operators.
The chain of
 (proper) inclusions $\rr{S}^{1}\subset\rr{S}^{1^+}_0\subset\rr{S}^{1^+}\subset\rr{S}^{1+\epsilon}$ holds true for every $\epsilon>0$.
To define a trace functional with domain the Dixmier ideal $\rr{S}^{1^+}$ we need to fix a \emph{generalized scale-invariant limit}\footnote{{
A generalized scale-invariant limit is a continuous positive linear functional $\omega: \ell^{\infty}(\mathbb{N}) \to \mathbb{C}$ which  coincides with the ordinary limit on the subspace of convergent
sequences and is invariant under \virg{dilations} of the sequences of the type $\{a_1,a_2,a_3,\ldots\}\mapsto \{a_1,a_1,a_2,a_2,a_3,a_3\ldots\}$}.}
$\omega: \ell^{\infty}(\mathbb{N}) \to \mathbb{C}$.
The $\omega$-{Dixmier trace} of a positive element of the Dixmier ideal is defined as
\[
  {\Tr}_{{\rm Dix},\omega}(T)\;: =\;  \omega( \{ \gamma_{N}(T)\}_{N} )
  \;,\qquad T \in \rr{S}^{1^+}\;,\;\; T\geqslant0\;.
\]
The definition of ${\Tr}_{{\rm Dix},\omega}$
extends to non-positive elements of  $\rr{S}^{1^+}$ by linearity. The
$\omega$-Dixmier trace
provides an example of a singular (hence non-normal) trace and it is continuous with respect to the  norm~\eqref{eq:clad_norm} \ie,
$|{\Tr}_{{\rm Dix},\omega}(T)|\leqslant \|T\|_{1^+}$. 
Every Dixmier trace fulfills the {cyclicity property}
$$
{\Tr}_{{\rm Dix},\omega}(TA)\;=\;{\Tr}_{{\rm Dix},\omega}(AT)\;,\qquad \forall\;\; T\in \rr{S}^{1^+}\;,\;\; A\in \bb{B}(\s{H})
$$
 the \emph{H\"older inequalities} \cite[Proposition 7.16]{gracia-varilly-figueroa-01} which, in the special case $p=1$ and $q=+\infty$, provide
 $$
 {\Tr}_{{\rm Dix},\omega}(|ATB|)\;\leqslant\;\|A\|\|B\|\; {\Tr}_{{\rm Dix},\omega}(|T|)\qquad\forall\;\; T\in \rr{S}^{1^+}\;,\;\; A,B\in \bb{B}(\s{H})\;.
 $$

 \medskip

 An element $T\in \rr{S}^{1^+}$ is called \emph{measurable} if  
the value of $\omega (\{\gamma_N(T)\}_N)$ is independent of the
  choice of the generalized scale-invariant limit $\omega$. For a positive element $T\geqslant 0$ this is equivalent to the convergence of a certain Ces\`{a}ro mean of $\gamma_N(T)$ Moreover, the set of measurable operators $\rr{S}^{1^+}_{\rm m}$ is
 a closed subspace of
$\rr{S}^{1^+}$ (but not an ideal) which is invariant under conjugation by bounded invertible operators \cite[Chap.~4, Sect.~2, Proposition 6]{connes-94}.
Evidently, $\rr{S}^{1^+}_0\subset\rr{S}^{1^+}_{\rm m}$.
A compact operator $T$ is called \emph{Tauberian} \cite[Definition 9.7.1]{lord-sukochev-zanin-12} if    the limit
  \[
 \lim_{N\to\infty}\left(\frac{1}{\log(N)}\sum_{n=0}^{N-1}\lambda_n(T)\right)\;=\; L\;,\]
exists. Here,   $\lambda_n(T)$ denotes an 
\emph{eigenvalue sequence} \cite[Definition 1.1.10]{lord-sukochev-zanin-12} of $T$. A non-negative operator $T\geqslant 0$ is 
  {Tauberian} if and only if $T\in\rr{S}^{1^+}_{\rm m}$  is measurable \cite[Theorem 9.3.1]{lord-sukochev-zanin-12}, and in that case
   \[
  {\Tr}_{{\rm Dix}}(T)\;: =\;
  \lim_{N\to\infty}\left(\frac{1}{\log(N)}\sum_{n=0}^{N-1}\mu_n(T)\right)\;,\]
  where the equality $\lambda_n(T)=\mu_n(T)$ has been used. However, not every element in $\rr{S}^{1^+}_{\rm m}$ is {Tauberian} as shown in \cite[Example 9.7.6]{lord-sukochev-zanin-12}.

\medskip

The following result provides a useful criterion to determine whether
 $\text{\upshape Tr}_{\text{\upshape Dix}}(T)$ is independent of $\omega$ and to compute its value.

\begin{theorem}[The Tauberian criterion]\label{theo:zeta_connes}
Let $T\geqslant 0$ be a non-negative compact operator such that $T^{1+x} \in \rr{S}^{1}$ 
for every $x>0$, and define the  \emph{zeta function}
$$
\zeta_{T}(x)\;:=\;\text{\upshape Tr}\left(T^{1+x}\right)\;.
$$ 
Then, the \emph{residue condition}
$$
\lim_{x\to0^+}\ x\;\zeta_{T}(x)\;=\;L
$$
implies that  $T\in \rr{S}^{1^+}_{\rm m}$ and
$$
\text{\upshape Tr}_{\text{\upshape Dix}}(T)\;=\;
\lim_{N\to\infty}\frac{1}{\log(N)}\sum_{n=0}^{N-1}\mu_n(T)\;=\;L\;,
$$
independently of the choice of $\omega$.
\end{theorem}
\proof[Proof (a sketch of)]
There are various proof of this result in the literature (see \eg \cite[Theorem 9.3.1]{lord-sukochev-zanin-12} or \cite[Section 1.4]{alberti-matthes-02}). For  the way the claim above is formulated we prefer to follows the strategy used in
\cite{gracia-varilly-figueroa-01}. Let $a_{k}:=\mu_{k-1}(T)$, with $n\in\N$, be the $n$-the singular value of $T$ and $s:=1+x$.
The condition 
$T^{1+x} \in \rr{S}^{1}$ 
for every $x>0$ translates in the fact that the series
$\sum_{k=1}^{+\infty}a_k^s$ is convergent for every $s>1$. The validity of the residue condition translates into
$$
\lim_{s\to1^+}\;(s-1)\sum_{k=1}^{+\infty}a_k^s\;=\;L\;
$$
and \cite[Lemma 7.20]{gracia-varilly-figueroa-01} provides
$$
\lim_{N\to+\infty}\frac{1}{\log\left(a_N^{-1}\right)}\sum_{k=1}^Na_k\;=\;L\;.
$$
The latter equation along with  \cite[Lemma 7.19]{gracia-varilly-figueroa-01} imply
$$
\lim_{N\to+\infty}\frac{1}{\log\left(N\right)}\sum_{k=1}^Na_k\;=\;\lim_{N\to+\infty}\frac{1}{\log\left(N\right)}\sum_{n=0}^{N-1}\mu_n(T)\;=\;L\;.
$$
The last equality automatically implies 
that $T\geqslant 0$ is Tauberian, and in turn
$T\in \rr{S}^{1^+}_{\rm m}$ with $\text{\upshape Tr}_{\text{\upshape Dix}}(T)=L$.
\qed

%------------------------------%


\begin{thebibliography} {[RMCPV]}
\frenchspacing \baselineskip=12 pt plus 1pt minus 1pt


\bibitem[AM]{alberti-matthes-02}  Alberti, P.~M.; Matthes, R.:
{\sl Connes' Trace Formula and Dirac Realization of Maxwell and Yang-Mills Action}. In: 
{\em Noncommutative Geometry and the Standard Model of Elementary Particle Physics}, Lecture Notes in Physics {\bf 596} 
(F. Scheck, W. Werner and H. Upmeier eds.). Springer, Berlin, 2002,  pp. 40-74



\bibitem[AMSZ]{azamov-mcdonald-sukochev-zanin-19} 
{Azamov, N.; McDonald, E.; Sukochev, F.; Zanin, D.:}
{\sl A Dixmier trace formula for the density of states}. 
Commun. Math. Phys.,  (2020)







\bibitem[CL]{carmona-lacroix-90}
{Carmona, R.; Lacroix, J.}: 
{\em Spectral Theory of Random Schr\"odinger Operators}. Birkh{\"{a}}user, Basel-Boston-Berlin, 1990 


\bibitem[CM]{connes-moscovici-95}  
Connes,  A.; Moscovici, H.:  
{\sl The local index formula in noncommutative geometry}.  
{Geom. Func. Anal.}~{\bf 5}, 174-243 (1995)




\bibitem[Con3]{connes-94}  Connes,  A.:  {\em Noncommutative Geometry}. Academic Press, San Diego, 1994


\bibitem[DGM]{denittis-gomi-moscolari-19} 
{De Nittis,~G.; Gomi, K.; Moscolari, M.:} 
 {\sl The geometry of (non-abelian) landau levels}. 	J. Geom. Phys.
{\bf 152}, 103649 (2020) 

\bibitem[DS]{denittis-sandoval-00} 
{De Nittis,~G.; Sandoval, M.:} 
{\sl The noncommutative geometry of the Landau Hamiltonian: Metric aspects}. 
SIGMA {\bf 16}, 146 (2020) 


\bibitem[Dix]{dixmier-81} 
{Dixmier, J.}: 
{\em Von Neumann Algebras}. 
North-Holland Publishing Co., Amsterdam, 1981




\bibitem[Fol2]{folland-99}  
{Folland, G. B.}:  
{\em Real Analysis: Modern Techniques and Their Applications}.
John Wiley {\&} Sons Inc., New York, 1999



\bibitem[GBVI]{gracia-varilly-88}  
{Gracia-Bondia, J. M., Varilly, J. C.}:  
{\sl Algebras of distributions suitable for phase-space quantum mechanics. I}.
J. Math. Phys. {\bf 29}, 569-623 (1988)


\bibitem[GBVII]{gracia-varilly-88-II}  
{Gracia-Bondia, J. M., Varilly, J. C.}:  
{\sl Algebras of distributions suitable for phase-space quantum mechanics. II. Topologies on the Moyal algebra}.
J. Math. Phys. {\bf 29}, 880-887 (1988)


\bibitem[GBVF]{gracia-varilly-figueroa-01}  Gracia-Bondia, J. M., Varilly, J. C.,  Figueroa,  H.:  {\em Elements of Noncommutative Geometry}.  Birkh\"{a}user, Boston, 2001




\bibitem[Gree]{greenleaf-69}  Greenleaf, F.~P.: {\em  Invariant Means on Topological Groups And Their Applications}.  Van Nostrand
Reinhold Co., New York, 1969





\bibitem[Hal]{halmos-74} 
{Halmos, P. R.}: 
{\em Measure Theory}. 
Springer-Verlag, New York,  1974




\bibitem[LLW]{loring-lu-watson-21}
{Loring T. A., Lu, J.; Watson, A. B.}:
{\sl Locality of the windowed local density of states}.
 \href{https://arxiv.org/abs/2101.00272}{preprint arXiv:2101.00272},  (2021)
 
\bibitem[LSZ]{lord-sukochev-zanin-12} 
Lord, S.; Sukochev, F.; Zanin, D.: 
{\em Singular Traces}. De Gruyter, Berlin,  2012





\bibitem[Nak]{nakamura-01} 
{Nakamura, S.}:
{\sl A remark on the Dirichlet-Neumann decoupling and the integrated density of states}. 
J. Funct. Anal. {\bf 179}, 136-152 (2001) 


\bibitem[OMS]{oldham-myland-spanier-09} 
{Oldham, K.; Myland, J.; Spanier, J.}: 
{\em An Atlas of Functions}. 
Springer,  2009




\bibitem[PF]{pastur-figotin-92} 
{Pastur, L.; Figotin, A.}: 
{\em Spectra of Random and Almost-Periodic Operators}. Academic Springer-Verlag, Berlin-Heidelberg-New York, 1992




\bibitem[RS]{reed-simon-I} 
{Reed,~M.; Simon,~B.}: 
{\em Methods of Mathematical Physics I: Functional Analysis}. Academic Press, Inc., San Diego, 1980





\bibitem[Sim]{simon-05} 
{Simon,~B.}: 
{\em Trace Ideals and Their Applications}. 
AMS, 2005




\bibitem[Ves]{veselic-08} Veseli\'{c}, I.: 
{\em Existence and Regularity Properties of the Integrated Density of States of Random Schr\"{o}dinger Operators}. Springer, Berlin-Heidelberg,  2008




\bibitem[Zak1]{zak1} Zak, J.: 
{\sl Magnetic translation groups}. 
{Phys. Rev. A}~{\bf 134}, 1602-1607 (1964)

\bibitem[Zak2]{zak2} Zak, J.: 
{\sl Magnetic translation groups II: Irreducible representations}. 
{Phys. Rev. A}~{\bf 134}, 1607-1611 (1964)





 \end{thebibliography}
\end{document}